\documentclass[
  aps,
  prb,
  reprint,
  superscriptaddress,
  longbibliography,
  nofootinbib
]{revtex4-2}

\usepackage{amsmath,amssymb,bm}
\usepackage{booktabs}
\usepackage{graphicx}
\usepackage{placeins}
\usepackage{xcolor}
\usepackage[colorlinks=true,allcolors=blue!55!black]{hyperref}
\usepackage{microtype}

\newcommand{\bPhi}{\bm\Phi}
\newcommand{\bk}{\bm k}
\newcommand{\br}{\bm r}
\newcommand{\av}[1]{\langle #1\rangle}

\begin{document}

\title{Vestigial chirality from fluctuating loop currents on the kagome lattice}

\author{Yin Shi}
\email{yin.shi@iphy.ac.cn}
\affiliation{Beijing National Laboratory for Condensed Matter Physics and Institute of Physics, Chinese Academy of Sciences, Beijing 100190, China}%
\author{Xuande Bu}
\affiliation{Beijing National Laboratory for Condensed Matter Physics and Institute of Physics, Chinese Academy of Sciences, Beijing 100190, China}%
\affiliation{University of Chinese Academy of Sciences, Beijing 100049, China}%
\author{Sheng Meng}
\email{smeng@iphy.ac.cn}
\affiliation{Beijing National Laboratory for Condensed Matter Physics and Institute of Physics, Chinese Academy of Sciences, Beijing 100190, China}%
\affiliation{University of Chinese Academy of Sciences, Beijing 100049, China}%
\affiliation{Songshan Lake Materials Laboratory, Dongguan, Guangdong 523808, China}%

\date{\today}

\begin{abstract}
Multicomponent order can melt in stages, leaving a composite order after its
primary constituents become short ranged. We study this possibility for
commensurate three-$Q$ loop-current order on the kagome lattice. Large-scale
cluster and parallel-tempering Monte Carlo simulations reveal direct and
two-stage melting regimes in an effective fixed-amplitude sign model.  In the latter regime,
translation-sector domain walls proliferate before chirality-changing walls, restoring
lattice translational symmetry while preserving long-range time-reversal-odd order. The
primary $M$-point correlations are short ranged in this intermediate phase.
The two-stage regime begins when the lowest-energy chirality-changing wall is
only about $12$--$13\%$ more costly than a same-chirality translation wall.
Finite-size scaling of a stable amplitude-resolved Ginzburg--Landau theory
shows that the phase survives amplitude relaxation. Our results establish a
quantitative domain-wall criterion for vestigial loop-current order and a
fluctuation route to time-reversal symmetry breaking without long-range
loop-current Bragg order. This separation provides a possible thermodynamic
framework for time-reversal-odd responses  
recently reported 
above the ordering temperature for conventional
charge density waves in kagome metals.
\end{abstract}

\maketitle

\section{Introduction}

The kagome metals $A\mathrm{V}_3\mathrm{Sb}_5$, where $A$ represents an alkali metal element, have been discovered as a family of
topological metals with superconducting ground states
\cite{Ortiz2019,Ortiz2020,Neupert2022}, combining a commensurate $2\times2$
charge-density-wave (CDW) transition, strong precursor fluctuations, and
reported signatures of broken time-reversal symmetry
\cite{Zhao2021,Jiang2021,Mielke2022,Chen2022,Subires2023,Graham2024}.
Microscopic studies connect the competing multi-$Q$ particle--hole channels to
van Hove singularities and sublattice interference
\cite{Denner2021,Park2021,LinNandkishore2021,Li2024,Zhan2026}.  Complementary
low-energy classifications and fluctuation-mediated theories show how
nonlocal interactions can stabilize charge-bond and loop-current orders
\cite{Feng2021,Christensen2022,Tazai2023}.  Here loop-current order denotes the
imaginary, time-reversal-odd part of a modulated electronic hopping.  The
candidate loop-current state contains three symmetry-related $M$-point
components and therefore breaks both translation and time-reversal symmetry.  This multicomponent structure poses a question about the finite-temperature phase structure that cannot be resolved by identifying the leading electronic instability alone: must the two symmetries be restored at the same transition upon heating?

Recent experiments make a separated thermal hierarchy especially timely.
Circular-dichroism angle-resolved photoemission on
$\mathrm{CsV}_3\mathrm{Sb}_5$ finds a time-reversal-odd response below
$T^*\simeq145$--$150\,\mathrm K$, above the conventional CDW transition near
$94\,\mathrm K$; it is interpreted as loop-current order that precedes the
inverse-Star-of-David real charge-bond reconstruction \cite{Cha2026}.  A
depth-resolved $\mu$SR study likewise reports a near-surface magnetic response
in $\mathrm{RbV}_3\mathrm{Sb}_5$ beginning near $175\,\mathrm K$, above the
bulk charge-order temperature near $110\,\mathrm K$ \cite{Graham2024}.
Here ``conventional CDW order'' denotes the observable real charge/bond and
lattice reconstruction; the relative weights of site charge, bond modulation,
and atomic displacement depend on the material and probe
\cite{Tan2021,Christensen2021}.  It should not be
identified automatically with the time-reversal-odd loop-current component.
These observations are therefore compatible not only with primary loop-current
order preceding the conventional CDW, but also with a vestigial chiral regime
in which a uniform time-reversal-odd composite remains ordered after the
primary translation-breaking loop-current fields become short ranged.  They do
not by themselves distinguish these possibilities: the loop-current model used
to interpret the photoemission data already doubles the unit cell size, and absence
of the real structural reconstruction does not establish restored translations
in the loop-current sector.

If the primary translation-breaking field disorders while a uniform composite
remains ordered, the intervening state is vestigial
\cite{Fernandes2019}.  Establishing such a phase requires more than observing a
nonzero composite moment on a finite lattice.  One must demonstrate long-range
order in a symmetry-nontrivial composite and short-ranged primary correlations
in the thermodynamic limit.  Tsvelik and Sarkar proposed a related phase-only
theory that identified a fluctuation-driven composite CDW crossover
\cite{Tsvelik2023}.  Subsequent Monte Carlo simulations of the compact
three-phase model found one continuous transition of the primary fields
together with that crossover \cite{Wildeboer2024}. The phase-locked composite
in these works is time-reversal even, so the crossover is not associated with any symmetry breaking. These
studies therefore do not establish a thermodynamic phase with long-range
time-reversal-odd order and short-ranged primary correlations.  In the strongly
commensurate limit studied here, the asymptotic phasons are pinned and melting
is instead controlled by domain walls \cite{McMillan1976}.

\begin{figure*}[t]
 \centering
 \includegraphics[width=0.98\textwidth]{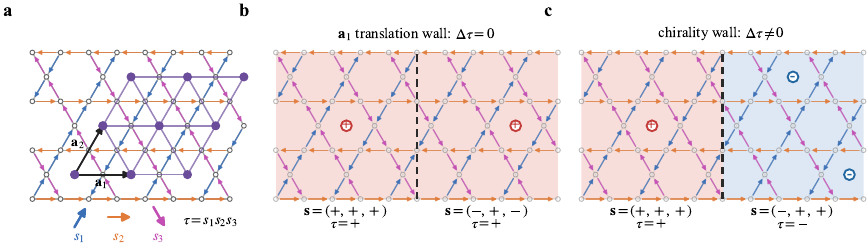}
 \caption{Coarse graining and domain-wall classes.  (a) Gray sites and bonds
 show the microscopic kagome lattice; purple sites at unit-cell centers and
 their links form the triangular GL regulator.  Each GL site carries the three
 internal $M$-point components $s_1,s_2,s_3$.  Colored arrows show their
 nearest-neighbor current form factors in the convention of
 Ref.~\cite{Zhan2026}; reversing $s_a$ reverses the corresponding extended
 current pattern.  The second-neighbor current channel is omitted for clarity.
 The symbols $\tau=\pm$ mark only the circulating-current hexagons of the
 $2\times2$ texture, separated by doubled Bravais vectors.  (b) An
 $\bm a_1$-translation wall connects $(+,+,+)$ and $(-,+,-)$: two signs
 reverse while
 $\tau=s_1s_2s_3$ is unchanged.  (c) The lowest-energy chirality-changing wall reverses one sign
 and therefore $\tau$.  The domain hue encodes chirality only.  The schematic
 Python script was developed with assistance from OpenAI Codex with GPT-5.6
 Sol; the graphical content and physical assignments were verified by the
 authors.}
 \label{fig:walls}
\end{figure*}

Low-energy and Landau model analyses have classified the relevant three-$Q$ charge,
bond, and flux orders \cite{Park2021,Feng2021,Christensen2021,Christensen2022};
the symmetry-complete homogeneous bond--flux theory is given in Refs.~\cite{Wagner2023,WagnerErratum2024}.
This classification selects candidate ordered
states, but homogeneous theory alone cannot decide how their spatially varying
configurations melt.  The essential energetic question is whether a domain wall that
reverses chirality can remain sufficiently more costly than a wall that changes
the translation domain while preserving chirality; see Fig.~\ref{fig:walls}.  

We answer this question in two stages.  First, we derive an effective fixed-amplitude sign model for the three-$Q$ loop current order, which 
allows cluster simulations through a linear system size $L=384$, maps the global phase topology,
and converts its multicritical endpoint into a ratio of microscopic wall
tensions.  Second, a stable amplitude-resolved lattice Ginzburg--Landau (GL) model
tests whether the two-stage sequence survives wall-core relaxation.  We find that a
vestigial chiral phase emerges when a chirality-changing wall is
only $12$--$13\%$ more costly than the same-chirality wall. This phase
also survives amplitude fluctuations in a controlled soft-spin parameter
family.  This is a sufficiency result for the phenomenological theory, not yet
a material-specific prediction: a microscopic calculation of the two relaxed
wall tensions remains necessary.

\section{Results}
\label{sec:fields}

\subsection{\texorpdfstring{Three-$Q$}{Three-Q} order and composite chirality}

The kagome Bravais lattice is triangular.  We place one coarse-grained field $\Phi$
on each kagome unit cell at
$\br_i=x_i\bm a_1+y_i\bm a_2$, where $\bm a_1$, $\bm a_2$ are the primitive 
lattice vectors and $x_i$, $y_i$ are integer coordinates.  Its three internal components
$\Phi_{ai}$, $a=1,2,3$, are real amplitudes of the symmetry-related $M$-point
loop-current patterns; they are not additional spatial sites.  Microscopic sublattice and bond
structure enters through the transformation laws and current form factors of
the components.  The triangular lattice is therefore a symmetry-compatible
coarse graining of kagome unit cells, not a replacement of the microscopic
kagome geometry.

With a convenient ordering at the $M$ points, primitive translations act as
\begin{equation}
\begin{aligned}
 T_{\bm a_1}:&\ (\Phi_1,\Phi_2,\Phi_3)
 \mapsto(-\Phi_1,+\Phi_2,-\Phi_3),\\
 T_{\bm a_2}:&\ (\Phi_1,\Phi_2,\Phi_3)
 \mapsto(+\Phi_1,-\Phi_2,-\Phi_3),
\end{aligned}\label{eq:translations}
\end{equation}
while a threefold rotation cycles the component labels.  Time reversal acts
as $\mathcal T:\bPhi\mapsto-\bPhi$.  Consequently,
\begin{equation}
 \chi_i=\Phi_{1i}\Phi_{2i}\Phi_{3i}
 \label{eq:chirality}
\end{equation}
is translation invariant and time-reversal odd.  Long-range order of $\chi$
with short-ranged correlations of every $\Phi_a$ therefore defines a uniform
vestigial chiral phase.  It preserves the translations restored by melting the
$M$-point order while continuing to break $\mathcal T$.

\subsection{Amplitude-resolved effective Hamiltonian}

Real bond and loop-current orders are naturally intertwined because they are
the real and imaginary parts of modulated electronic hopping.  They generally
transform as different irreducible representations, however, and need not be
simultaneously critical.  Denoting the bond field by $\bm B$, the
bond and loop-current combination permits the cubic coupling
$-g(B_1\Phi_2\Phi_3+B_2\Phi_3\Phi_1+B_3\Phi_1\Phi_2)$
\cite{Wagner2023,WagnerErratum2024}.  We assume that the bond field is massive
on the scales of interest and integrate it out rather than retain it as an
independent Monte Carlo field.

To see the resulting terms, define $X_{1i}=\Phi_{2i}\Phi_{3i}$ and its cyclic
permutations.  If $\mathcal A_B$ is the positive quadratic kernel of the
massive bond field, its leading contribution to the Hamiltonian is
\begin{equation}
 H_B=\frac12\sum_{ij,a}B_{ai}(\mathcal A_B)_{ij}B_{aj}
 -g\sum_{i,a}B_{ai}X_{ai}+\cdots,
\end{equation}
and integrating out $\bm B$ generates
\begin{equation}
 \Delta H_\Phi=-\frac{g^2}{2}\sum_{ij,a}
 X_{ai}(\mathcal A_B^{-1})_{ij}X_{aj}+\cdots.
 \label{eq:integrate-B}
\end{equation}
In the local massive limit,
$\mathcal A_B^{-1}\simeq r_B^{-1}\delta_{ij}$, where $r_B>0$ is the local
bond-field mass, Eq.~\eqref{eq:integrate-B}
becomes
$-g^2(2r_B)^{-1}\sum_{i,a<b}\Phi_{ai}^2\Phi_{bi}^2$ and is absorbed into the
quartic coefficients below.  The allowed bond-field cubic invariant generates
a term proportional to $\chi_i\chi_j$ at higher order.  Thus
eliminating a noncritical bond field preserves the loop-current symmetries and
motivates, but does not microscopically determine, the effective couplings.

Our production Hamiltonian is consequently the three-field theory
\begin{equation}
\begin{aligned}
 H_{\rm GL}={}&\frac{K_\Phi}{2}\sum_{\langle ij\rangle}
 |\bPhi_i-\bPhi_j|^2+\sum_i\left[
 \frac{r_\Phi}{2}\Phi_i^2+\frac{u_\Phi}{4}(\Phi_i^2)^2\right.\\
 &\left.-\frac{v_\Phi}{2}\sum_{a<b}\Phi_{ai}^2\Phi_{bi}^2
 +\frac{c_\Phi}{6}(\Phi_i^2)^3\right]-J_\chi\sum_{\langle ij\rangle}\chi_i\chi_j.
\end{aligned}\label{eq:HGL}
\end{equation}
Here $\Phi_i^2=\sum_a\Phi_{ai}^2$, and the coefficients already include the
short-distance renormalizations generated by the eliminated bond field.
The interaction proportional to $J_\chi$ is sixth order in the primary fields
and allowed by every symmetry.  It directly changes the relative cost of
chirality-changing and translation-related domain walls (Fig.~\ref{fig:walls}).  
We interpret it as an
effective interaction generated after shorter-scale electronic, magnetic, or
collective modes are integrated out, rather than as a coefficient determined
by the homogeneous quartic theory.  The on-site sextic term bounds the
soft-spin energy.  For the triangular bond counting used here, a sufficient
strict condition is $c_\Phi\geq0,\quad |J_\chi|<3c_\Phi/2$.

\subsection{Fixed-amplitude limit and domain wall energetics}
\label{sec:sign}

Deep in an equal-amplitude three-$Q$ state we write
$\Phi_{ai}=A s_{ai}$ with $s_{ai}=\pm1$.  Up to constants,
Eq.~\eqref{eq:HGL} reduces to
\begin{equation}
 H_{\rm sign}=-K_1\sum_{\langle ij\rangle,a}s_{ai}s_{aj}
 -K_3\sum_{\langle ij\rangle}\tau_i\tau_j,
 \quad \tau_i=s_{1i}s_{2i}s_{3i},
 \label{eq:Hsign}
\end{equation}
with
\begin{equation}
 K_1=K_\Phi A^2,\quad K_3=J_\chi A^6,\quad
 q\equiv\frac{K_3}{K_1}=\frac{J_\chi A^4}{K_\Phi}.
 \label{eq:q}
\end{equation}
We denote by $T_s$ the transition at which the primary component signs lose
long-range order upon heating and by $T_\tau$ the transition at which the
composite chirality disorders.
The primary phase orders both $s_a$ and $\tau$, and hence breaks translations
and time reversal. In the possible vestigial phase, $\tau$ remains ordered while
$\av{s_a(\br)s_a(0)}$ is short ranged for every component.

The limiting cases provide exact checks.  At $q=0$, the three components are
independent ferromagnetic Ising models on a triangular lattice, with a critical temperature $T_c(q=0)=4K_1/\ln3$~\cite{Houtappel1950}.
For $q\to\infty$, $\tau$ orders at the triangular-Ising scale 
\begin{equation}
 T_\tau(q\to\infty)=\frac{4K_3}{\ln3}.
 \label{eq:ising-limit}
\end{equation}
Within a fixed-$\tau$ sector the four allowed component
states form a ferromagnetic four-state Potts model whose bond contrast is
$4K_1$, giving a separate critical temperature~\cite{Wu1982}
\begin{equation}
 T_s(q\to\infty)=\frac{4K_1}{\ln2}.
 \label{eq:potts-limit}
\end{equation}

\begin{figure*}[t]
 \centering
 \includegraphics[width=0.98\textwidth]{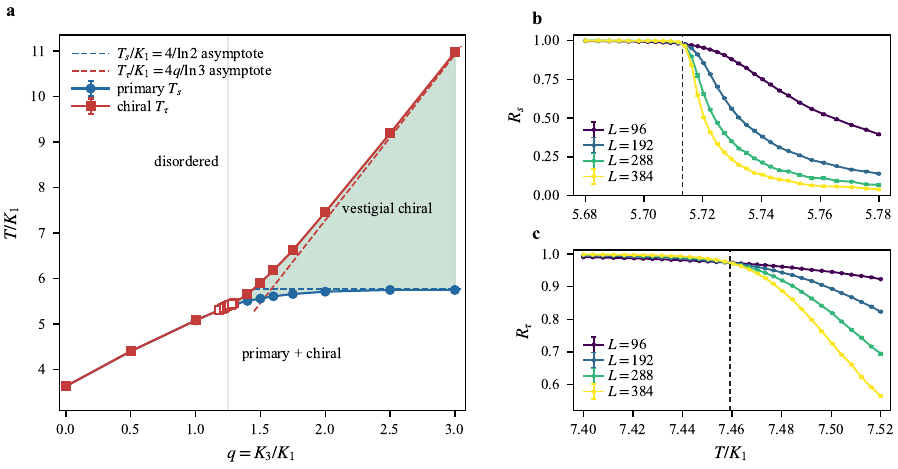}
 \caption{Phase diagram and representative crossings of the sign model.
 (a) Primary (blue circles) and chiral (red squares) transition temperatures.
 Open symbols denote transport-limited estimates whose errors include the
 temperature-grid floor.  Green shading marks the vestigial phase and the gray
 strip the conservative endpoint corridor
 $1.24\lesssim q_c\lesssim1.26$; dashed curves are the exact asymptotes.
 (b,c) Correlation ratios $R_s$ and $R_\tau$ for
 $L=96,192,288,384$ at $q=2$.  Vertical dashed lines mark the joint-bootstrap
 $L=288$--$384$ crossings, $T_s/K_1=5.7131(7)$ and
 $T_\tau/K_1=7.4592(8)$.  Points are measured temperatures, lines guide the
 eye, and error bars are joint chain/block-bootstrap standard errors.  The
 plotting Python script was developed with assistance from OpenAI Codex with
 GPT-5.6 Sol; the underlying data and rendered figure were verified by the
 authors.}
\label{fig:sign-phase}
\end{figure*}

The sign limit turns the phase-topology threshold into a wall-energy target; 
see Fig.~\ref{fig:walls}.
A straight same-chirality translation wall flips two component signs.  If it
cuts $\nu_{\hat n}$ triangular bonds per unit length, its tension is
\begin{equation}
 \sigma_{\rm even}=4\nu_{\hat n}K_1.
 \label{eq:even-wall}
\end{equation}
The cheapest chirality-changing wall flips one component and has tension
\begin{equation}
 \sigma_{\rm odd}=2\nu_{\hat n}(K_1+K_3).
 \label{eq:odd-wall}
\end{equation}
For equal orientations,
\begin{equation}
 \frac{K_3}{K_1}=2\frac{\sigma_{\rm odd}}{\sigma_{\rm even}}-1.
 \label{eq:wall-map}
\end{equation}
Thus the numerical endpoint found below,
$1.24\lesssim K_3/K_1\lesssim1.26$, corresponds to
$\sigma_{\rm odd}/\sigma_{\rm even}\simeq1.12$--$1.13$.  Although the
coupling ratio is of order unity, the required wall-tension hierarchy is
modest because an odd wall already pays half of the two-component
translation-wall cost at $K_3=0$.

We determine the finite-temperature behavior using large-scale cluster and
parallel-tempering Monte Carlo simulations on periodic $L\times L$ triangular
lattices with $N=L^2$ sites.
The sign model is simulated through $L=384$, while the amplitude-resolved
validation reaches $L=192$.  Phase boundaries are located from crossings of
dimensionless finite-size ratios; structure-factor scaling and real-space
correlations independently test whether the primary fields are short ranged.
The update schemes, observable definitions, and uncertainty analysis are given
in Appendix Secs.~\ref{sec:methods-sampling}--\ref{sec:methods-statistics};
run-specific diagnostics and complete transition sequences are collected in
the Supplementary Information.

\subsection{Fixed-amplitude phase diagram}
\label{sec:sign-results}

Figure~\ref{fig:sign-phase}(a) summarizes the phase diagram obtained from
lattices as large as $L=384$.  At small $q$, the primary and composite orders
disappear together.  At large $q$, heating first disorders the component
signs at $T_s$ while preserving $\av{\tau}\neq0$; time reversal is restored
only at a second transition $T_\tau$.  The boundaries approach the exact Ising and Potts limits in Eqs.~\eqref{eq:ising-limit} and
\eqref{eq:potts-limit}.

At $q=2$, the crossings converge to
$T_s/K_1=5.7131(7)$ and $T_\tau/K_1=7.4592(8)$
[Fig.~\ref{fig:sign-phase}(b,c)].  Their wide separation establishes a robust
vestigial interval rather than a finite-size shoulder.  Binder ratios provide
an independent consistency check, while the predominantly single-peaked
energy distributions show no stable coexistence signal at either boundary.
The finite-size estimators are defined in
Appendix Sec.~\ref{sec:methods-observables}, their joint-bootstrap analysis is described
in Appendix Sec.~\ref{sec:methods-statistics}, and the full $q=2$ production
record and Binder curves appear in Supplementary Secs.~II A and IV.

Near the point where the two boundaries separate, the splitting is unresolved
at $q=1.24$ and first resolved at $q=1.26$.  We therefore obtain the
conservative endpoint corridor
\begin{equation}
 1.24\lesssim q_c\lesssim1.26.
 \label{eq:qc}
\end{equation}
The split then widens rapidly with increasing $q$.  Resolving the order of the direct
transition is a separate question and does not affect this phase topology.
The full phase-boundary estimates are tabulated in Supplementary Table~S1;
the endpoint criterion, tempering diagnostics, and histogram checks are given
in Supplementary Sec.~II B.

\subsection{Thermal and defect signatures}

\begin{figure*}[t]
 \centering
 \includegraphics[width=0.96\textwidth]{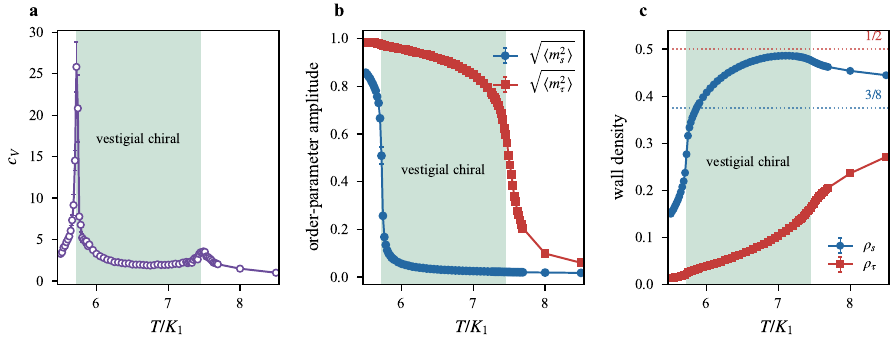}
 \caption{Thermal observables at $q=2$ and $L=96$: (a) per-site specific heat, (b) primary and chiral root-mean-square order parameters, and
 (c) primary-sector wall density $\rho_s$ and chiral-sector wall density
 $\rho_\tau$.  The former counts even-parity walls that disorder the $s_a$
 fields while preserving $\tau$; the latter counts walls that reverse $\tau$.
 Dotted lines are the random-sign limits $3/8$ and $1/2$.  Green shading marks
 the vestigial interval determined independently from the finite-size
 crossings in Fig.~\ref{fig:sign-phase}.  The root-mean-square order parameters and wall
 densities are defined in Appendix Sec.~\ref{sec:methods-observables}.  The
 plotting Python script was developed with assistance from OpenAI Codex with
 GPT-5.6 Sol; the underlying data and rendered figure were verified by the
 authors.}
 \label{fig:temperature}
\end{figure*}

\begin{figure*}[t]
 \centering
 \includegraphics[width=0.98\textwidth]{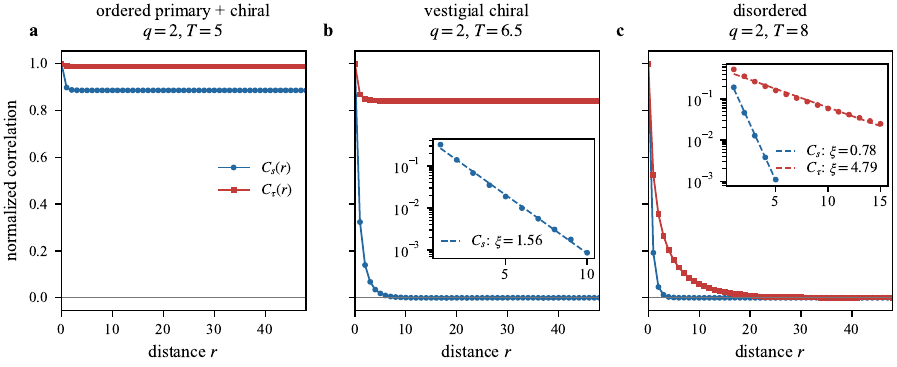}
 \caption{Direction-averaged correlations at $q=2$ and $L=96$ of the $s_a$ [$C_s(r)$] 
 and $\tau$ [$C_\tau(r)$] fields.  (a) Both
 $C_s(r)$ and $C_\tau(r)$ plateau in the primary chiral phase.  (b) In the
 vestigial phase, $C_s(r)$ decays within a few lattice spacings while
 $C_\tau(r)$ remains long ranged.  (c) Both channels decay above $T_\tau$.
 Every inequivalent integer separation is shown.  Semilog insets fit only the
 resolved short-range channels, where $\xi$ is the descriptive correlation
 length.  The plotting Python script was developed with assistance from OpenAI
 Codex with GPT-5.6 Sol; the underlying data and rendered figure were verified
 by the authors.}
\label{fig:correlations}
\end{figure*}

The specific heat $c_V$ in Fig.~\ref{fig:temperature}(a) exhibits a dominant sharp
peak at $T_s$ and a much weaker anomaly near $T_\tau$.  The lower feature
accompanies the loss of the three component orders [Fig.~\ref{fig:temperature}(b)].  
The upper feature marks the loss of the composite order and the restoration of time reversal.  
Its small amplitude explains why a single-size specific-heat curve does not by
itself resolve the two transitions reliably; we therefore locate both
boundaries from finite-size correlation-ratio crossings.  
The high-temperature tails in Fig.~\ref{fig:temperature}(b) decrease with
system size as expected for a disordered two-dimensional phase and therefore
represent finite-size susceptibility backgrounds, not residual order.
The energy-fluctuation definition of $c_V$ is given in Appendix Sec.~\ref{sec:methods-observables},
and the finite-size tail analysis is given in Supplementary Sec.~II C.

The domain wall densities in Fig.~\ref{fig:temperature}(c) reveal the underlying melting mechanism.  
At $T_s$, the translation wall density $\rho_s$ rises rapidly while the chirality wall density 
$\rho_\tau$ remains dilute.  The latter
grows strongly only near $T_\tau$.  The intermediate phase is therefore a
plasma of same-chirality translation walls in which time-reversal-changing
walls remain suppressed.  The wall observables are defined in
Appendix Sec.~\ref{sec:methods-observables}.

\begin{figure*}[t]
 \centering
 \includegraphics[width=0.96\textwidth]{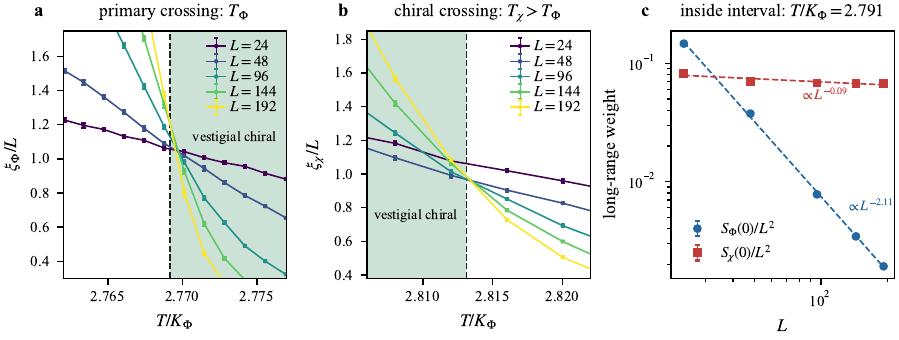}
 \caption{Amplitude-resolved vestigial order at $J_\chi/K_\Phi=5.5$.
 (a,b) Primary
 and chiral correlation-length ratios for $L=24,48,96,144,192$ cross at distinct
 temperatures.  Dashed lines mark the joint-block-bootstrap $L=144$--$192$
 estimates of the primary and chiral transition temperatures,
 $T_\Phi/K_\Phi=2.76918(25)$ and $T_\chi/K_\Phi=2.81310(64)$; green shading is the
 intervening window.  (c) Normalized zero-momentum structure factors inside
 the window at $T/K_\Phi=2.79121$; dashed lines are descriptive power-law fits.
 Here $S_X(0)$ is the spatially integrated two-point
 correlation in channel $X\in\{\Phi,\chi\}$ at zero momentum, and division by
 $L^2$ gives its intensive order-parameter normalization; see
 Eq.~\eqref{eq:structure-factor}.  The plotting Python script was developed
 with assistance from OpenAI Codex with GPT-5.6 Sol; the underlying data and
 rendered figure were verified by the authors.}
 \label{fig:soft-validation}
\end{figure*}

\subsection{Real-space characterization}

Real-space correlations provide the most direct distinction among the three
phases.  Figure~\ref{fig:correlations} uses the broad $q=2$ window so that the
narrow endpoint corridor does not obscure the limiting behaviors.  Both the $s$
and $\tau$ correlation functions $C_s(r)$ and $C_\tau(r)$ approach plateaus in the 
fully ordered phase, only $C_\tau(r)$
plateaus in the vestigial phase, and both are short ranged above $T_\tau$.
The correlation estimator is defined in Appendix Sec.~\ref{sec:methods-observables},
and its sampling details are reported in
Supplementary Sec.~II A.

\subsection{Robustness to amplitude fluctuations}
\label{sec:soft-results}

The sign model freezes $|\Phi_a|$.  We therefore restore local amplitude
fluctuations in Eq.~\eqref{eq:HGL} and follow a stable, relatively stiff
soft-spin family from the direct-transition regime into the two-transition
regime.  This family is designed as a controlled bridge from the sign limit,
not as a material-specific fit.  The parameter family and stability checks are
given in Supplementary Sec.~III A.  The update scheme is described in
Appendix Sec.~\ref{sec:methods-sampling}, while the production campaign and convergence
diagnostics are reported in Supplementary Sec.~III B.

The $J_\chi/K_\Phi=4.75$ case lies on the direct-transition side, while the
small separation at $J_\chi/K_\Phi=5.0$ remains unresolved.  At
$J_\chi/K_\Phi=5.5$, the largest size pair gives
$T_\Phi/K_\Phi=2.76918(25)$, $T_\chi/K_\Phi=2.81310(64)$, and
$(T_\chi-T_\Phi)/K_\Phi=0.04392(65)$
[Fig.~\ref{fig:soft-validation}(a,b)].  Agreement among successive large sizes
shows that the interval survives amplitude relaxation.

Separated crossings identify two thermal scales but do not by themselves
determine the order between them.  Figure~\ref{fig:soft-validation}(c) supplies
that test, depicting the normalized zero-momentum structure factors
$S_X(0)/L^2$ for the two fields $X\in\{\Phi,\chi\}$.  At
$T/K_\Phi=2.79121$ inside the interval,
$S_\Phi(0)/L^2$ approaches its short-range $L^{-2}$ scaling with increasing
size, whereas $S_\chi(0)/L^2$ approaches a nonzero plateau.  The
amplitude-resolved model therefore has short-ranged
primary loop-current correlations but long-range composite chirality.  The
second-moment estimator and phase-assignment tests are defined in
Appendix Sec.~\ref{sec:methods-observables}; the complete crossing sequence and scaling
values are given in Supplementary Tables~S2--S3 and Sec.~III C, with the
uncertainty procedure described in Appendix Sec.~\ref{sec:methods-statistics}.

\subsection{Connection to the fixed-amplitude regime}

Near the transitions, the measured root-mean-square component amplitude is
$A_{\rm rms}\simeq0.714$.  Substitution into Eq.~\eqref{eq:q}
gives $q_\chi^{\rm eff}\simeq(1.23,1.30,1.43)$ for
$J_\chi/K_\Phi=(4.75,5.0,5.5)$.  The first and last values lie below and above
the rigid-amplitude endpoint $q_c\simeq1.25$, consistent with their direct and
two-transition behaviors.  The middle value maps above the endpoint although
its thermodynamic splitting remains unresolved, showing that the bulk estimate
is not a quantitative threshold.

Wall-core relaxation explains the discrepancy: local suppression of $A$
weakens the chiral contribution more rapidly than the primary stiffness,
making chirality reversal easier than the uniform-amplitude estimate suggests.
The soft-spin onset is consequently shifted to larger bulk
$q_\chi^{\rm eff}$, and fully relaxed even- and odd-wall tensions are the
appropriate matching variables.  Numerical inputs are given in Supplementary
Sec.~III D.

\FloatBarrier
\section{Discussion}
\label{sec:discussion}

\subsection{Symmetry distinction and material criterion}

The sign and soft-spin models establish two-stage melting for their specified
classical Hamiltonians.  Although the numerical location of the endpoint is
nonuniversal, the order distinction is exact: the intermediate phase restores
the translations broken by the $M$-point fields while retaining a uniform
Ising variable that breaks time reversal.  The consistent correlation-ratio,
structure-factor, and real-space tests distinguish this state from a crossover
with a large but finite chiral correlation length.

The attractive $J_\chi$ interaction is a symmetry-allowed phenomenological
extension.  Because the local model with $J_\chi/K_\Phi=0$ did not resolve the same
phase, our result does not imply that it occurs in every kagome loop-current
system.  The wall mapping in Eq.~\eqref{eq:wall-map} states the material
requirement more directly than the bare coupling: the lowest-energy odd wall must be
about $12$--$13\%$ more costly than the lowest-energy even wall in the rigid limit.
Amplitude relaxation changes the numerical threshold but not the matching
principle.

A concurrent preprint by Iqbal proposed, on symmetry grounds, that vestigial chiral order is allowed in a doped kagome Dirac spin liquid~\cite{Iqbal2026}. 
This microscopic symmetry analysis is complementary to the present work, 
which establishes a separate finite-temperature vestigial chiral phase in a commensurate three-\(Q\) loop-current model through finite-size scaling and 
relates its stability to the relative tensions of chirality and translation domain walls.

\subsection{Microscopic routes to enhanced chirality-wall tension}

The required hierarchy is modest enough to have several plausible microscopic
origins.  First, an odd wall interpolates between domains with opposite orbital
magnetization and Berry curvature.  Its fermionic reconstruction may suppress
the particle--hole gap, redistribute more low-energy spectral weight, or bind
wall states, making it more costly than a translation wall connecting domains
with the same chirality.  The difference can be amplified near the intertwined
van Hove singularities that favor loop-current order
\cite{Li2024,Zhan2026}.  Wall-state entropy can partly offset this energetic
penalty, however, so the relevant comparison is between finite-temperature
free energies rather than zero-temperature band energies alone.

A second route is coupling to a soft time-reversal-odd collective mode, a
possibility consistent with fluctuation-mediated loop-current mechanisms
\cite{Tazai2023}.  For
an orbital-magnetization, scalar-spin-chirality, or magnetic field $M_i$ with
its Hamiltonian
\begin{equation}
 H_M=\sum_i\left(\frac{r_M}{2}M_i^2-\lambda M_i\chi_i\right)
 -J_M\sum_{\langle ij\rangle}M_iM_j,
 \label{eq:auxiliary-chiral-mode}
\end{equation}
where $r_M>0$ is the inverse local susceptibility, $J_M$ couples neighboring
auxiliary modes, and $\lambda$ is the local $M\chi$ coupling,
integrating out $M$ generates, to leading order, an attractive chirality
interaction $J_\chi^{\rm ind}\sim J_M\lambda^2/r_M^2$.  It is enhanced when
the auxiliary mode is close to its own instability and directly raises the
cost of an odd wall.  Microscopic current conservation and the sharing of
kagome bonds can instead bind elementary component walls into even-parity
pairs: an isolated one-component reversal is then penalized while a
two-component translation wall remains comparatively inexpensive.  Conversely,
sublattice interference, frustrated gradient terms, or proximity to a
competing ordering wavevector can soften the primary stiffness and lower the
even-wall tension without removing the chirality penalty.  Multicomponent ring
processes around kagome triangles or hexagons provide another possible source
of an effective chirality interaction, especially when flat bands or small
strong-coupling denominators enhance higher-order virtual processes.

These mechanisms show why $\sigma_{\rm odd}>\sigma_{\rm even}$ is physically
plausible, but they do not establish it for a particular compound.  A static
phonon or ordinary strain is time-reversal even and cannot align $\chi$
linearly, although it can modify amplitudes and primary stiffnesses.  Likewise,
a homogeneous sixth-order coefficient, or an estimate of $J_\chi$ alone,
cannot capture the different wall-core relaxations.  A material-specific test
must compute both fully relaxed wall free energies in the same electronic
theory and compare their ratio with the $1.12$--$1.13$ target.

\subsection{Relation to compact-phase composite order}

Refs.~\cite{Tsvelik2023,Wildeboer2024} study compact three-phase CDW models,
including commensuration effects, in which continuous phase fluctuations and
vortices govern melting.  Their translation-invariant composite is
time-reversal even in the proposed intermediate regime, so its development is
a crossover rather than a separate symmetry-breaking transition.  The Monte
Carlo study of Ref.~\cite{Wildeboer2024} accordingly found one thermodynamic
transition and did not establish long-range time-reversal-odd composite order
with short-ranged primary correlations.

By contrast, in our strongly commensurate model, the real loop-current
amplitudes have discrete translation domains.  Their product
$\chi=\Phi_1\Phi_2\Phi_3$ is explicitly odd under time reversal, so
$\langle\chi\rangle\ne0$ breaks an exact $\mathbb Z_2$ symmetry even when the
primary $M$-point correlations are short ranged.  The distinction is therefore
the domain-wall mechanism and a separate thermodynamic vestigial chiral phase,
established here by correlation-ratio crossings, structure-factor scaling,
and real-space correlations.

\subsection{Experimental implications}

Uniform time-reversal-odd order permits a zero-field anomalous Hall response,
Kerr rotation, and weak orbital magnetic fields, with their magnitudes set by
the coupling of $\chi$ to the electronic bands
\cite{Yang2020,Jiang2021,Mielke2022,Xu2022,Guo2022,Li2024,Zhan2026}.
Simultaneously, the primary $2\times2$
loop-current Bragg correlations are short ranged.  The characteristic thermal
sequence is therefore a time-reversal-sensitive onset at $T_\chi$, followed
only at lower temperature by sharp $M$-point order at $T_\Phi$.  Diffuse
$M$-point scattering and an enhanced chiral susceptibility should occupy the
intermediate window.  Opposite-$\chi$ domains can suppress bulk zero-field
signals, making field training, hysteresis, and local magnetic probes important
diagnostics.

Two recent observations are particularly suggestive of such a hierarchy.
The circular-dichroism photoemission response of
$\mathrm{CsV}_3\mathrm{Sb}_5$ onsets around $145$--$150\,\mathrm K$, well
above the conventional CDW transition, and was attributed to loop-current
order \cite{Cha2026}.  Near-surface $\mu$SR in
$\mathrm{RbV}_3\mathrm{Sb}_5$ similarly finds a static magnetic response above
the long-range charge-order temperature \cite{Graham2024}.  Diffuse x-ray
scattering also establishes quasi-static, short-range CDW precursors above
$T_{\rm CDW}$ \cite{Chen2022,Subires2023}.  Taken together, these results are
compatible with a time-reversal-odd regime preceding long-range real
charge-bond/lattice order.  They are not yet a thermodynamic identification of
the phase found here: structural CDW scattering does not directly measure the
$M$-point loop-current correlator, and the photoemission interpretation permits
primary translation-breaking loop-current order rather than only a uniform
composite.

The experimental status of time-reversal breaking is also not uniform across
probes.  Kerr, chiral-transport, and $\mu$SR studies have reported
time-reversal-odd responses \cite{Mielke2022,Xu2022,Guo2022,Graham2024},
whereas high-resolution zero-area-loop Sagnac measurements found no
spontaneous polar Kerr signal in $\mathrm{CsV}_3\mathrm{Sb}_5$ within a
$30\,$nrad noise floor \cite{Saykin2023}.
A decisive test of vestigial chirality would therefore combine a local or
domain-resolved time-reversal-sensitive probe with a wavevector-resolved probe
of the primary \emph{loop-current} correlations in the same sample and depth.
The required intermediate regime has long-range $\chi$ order but no
resolution-limited primary $M$-point peak.

These are symmetry-allowed signatures, not a prediction of quantized Hall
transport.  Whether vestigial order produces a Chern metal or insulator depends
on the microscopic coupling between $\chi$ and the fermions.  In particular,
long-range order of the composite does not make the one-electron Hamiltonian
equivalent to a static three-$Q$ texture with long-range $\Phi_a$ order.

\subsection{Relation to vestigial superconductivity and outlook}

The symmetry logic parallels charge-$4e$ and charge-$6e$ vestigial
superconductors: a composite remains coherent after its root order disorders
\cite{Agterberg2008,Berg2009,FernandesFu2021,Jian2021,Lin2025,Babaev2004,Herland2010}.  Let $\theta$
be the compact phase of an elementary
charge-$2e$ pair field, $\Delta_{2e}\propto e^{i\theta}$.  Its $n$th-power
composite $\Delta_{2ne}\sim(\Delta_{2e})^n\propto e^{in\theta}$ carries
charge $2ne$: $n=2$ and $3$ give charge-$4e$ and charge-$6e$ order, respectively.  In a
vestigial phase, correlations of $e^{i\theta}$ are short ranged while
$e^{in\theta}$ retains phase coherence.  Because $\theta$ is compact, the
defect spectrum can contain $2\pi/n$ fractional vortices, often tied to
discrete defects, whose binding and proliferation help control the transitions.

Our defect mechanism is different.  Strong commensuration locks the phasons,
leaves real component fields, and replaces the compact-phase defects by even-
and odd-parity domain walls.  The wall-tension ratio, rather than a
fractional-vortex fugacity, therefore controls the intermediate phase.  The
triangular lattice used here is the Bravais regulator
of kagome unit cells and should not be confused with the emergent kagome lattice
of phase variables in the charge-$6e$ model of Ref.~\cite{Lin2025}.

We have shown that commensurate three-$Q$ loop-current order can melt in two
stages.  In the sign model, a direct line gives way near
$K_3/K_1\simeq1.25$ to a phase that restores translations before time reversal.
This endpoint requires only a $12$--$13\%$ hierarchy between the lowest
chirality-changing and same-chirality wall tensions.  A stable five-size
soft-spin calculation retains a nonzero vestigial interval after amplitude
relaxation, and structure-factor scaling verifies that the primary loop-current
field is short ranged within it.  The remaining material question is therefore
concrete: determine the two relaxed wall tensions microscopically or
experimentally and compare their ratio with the criterion established here.

\section*{Note added}

After this manuscript was posted on arXiv, we became aware of a concurrent
preprint by Iqbal~\cite{Iqbal2026}, posted shortly before ours, which proposed
on symmetry grounds that vestigial chiral order is allowed in a doped kagome
Dirac spin liquid.  That work addresses a distinct microscopic setting and
symmetry classification and is complementary to our finite-temperature
demonstration of a thermodynamic vestigial chiral phase and its domain-wall
criterion.

\begin{acknowledgments}
OpenAI Codex with GPT-5.6 Sol assisted with manuscript organization and
language revision.  The authors directed its use, checked the cited literature
and scientific content, and reviewed and edited all generated text.  The
authors take full responsibility for the manuscript.
This work was supported by the start-up grant from the Institute of Physics, Chinese Academy of Sciences, and CAS Project for Young Scientists in Basic Research (Grant No. YSBR143).
S.M. acknowledges financial support from Ministry of Science and Technology (Grant No. 2021YFA1400200), National Natural Science Fund of China (Grants No.12450401 and No. 12025407), and Chinese Academy of Sciences (No.YSBR047).
\end{acknowledgments}

\section*{Data availability}

The numerical data supporting the findings of this study are available from the corresponding author upon reasonable request. The simulation inputs required to regenerate the results are included with the archived code described below.
The custom code used for the Monte Carlo simulations, data analysis, and figure generation, together with the production input configurations, is openly available under the MIT License on Zenodo~\cite{shi_2026_23254801}.

%

\appendix*

\section{Methods}
\label{sec:methods}

\subsection{Monte Carlo sampling}
\label{sec:methods-sampling}

All simulations use periodic $L\times L$ triangular Bravais lattices with
$N=L^2$ sites.  Angular brackets $\av{\cdots}$ denote thermal Monte Carlo
averages.  We sample the sign model with embedded Swendsen--Wang cluster updates
\cite{SwendsenWang1987}.  Fixed-temperature cluster scans map the broad phase
diagram; parallel tempering \cite{HukushimaNemoto1996} is used in the narrow
direct-transition corridor.  For each embedded Swendsen--Wang sub-update, the production
GPU implementation batches all coupling points, independent chains, and temperature 
replicas, and processes their lattice bonds or sites in parallel.  We benchmark this 
cluster updating algorithm against 
an independent local-update baseline.  Writing $\br=x\bm a_1+y\bm a_2$, we assign 
colors to sites by $c(x,y)=(x-y)\bmod3\in\{0,1,2\}$.  Every nearest-neighbor move 
changes $c$ by $\pm1$ modulo three, so no bond connects equal colors.  Sites of one 
color can therefore be updated concurrently with their neighbors fixed; cycling through
$c=0,1,2$ gives one conflict-free GPU local sweep.

The soft-spin model combines additive component proposals with two exact
embedded-cluster updates.  For a one-component reflection of component $a$,
the amplitudes and the other two sign fields define a signed conditional Ising
model with bond
\begin{equation}
 J_{ij}^{(a)}=K_\Phi|\Phi_{a,i}\Phi_{a,j}|
 +J_\chi|\chi_i\chi_j|
 \prod_{b\ne a}\operatorname{sgn}(\Phi_{b,i}\Phi_{b,j}).
 \label{eq:single-cluster-bond}
\end{equation}
Satisfied bonds of magnitude $|J_{ij}^{(a)}|$ are activated with probability
$1-\exp[-2|J_{ij}^{(a)}|/T]$.  This update changes $\chi$ and samples odd
walls.  A second cluster reflection changes two components together.  Its
conditional bond is
$J_{ij}^{(ab)}=K_\Phi(\Phi_{a,i}\Phi_{a,j}+\Phi_{b,i}\Phi_{b,j})$ because every
onsite invariant and $\chi_i$ are unchanged by the paired reflection.  It
therefore moves the primary translation walls without an intermediate chirality
wall.  Both clusters are needed: at $L=96$, adding the paired update reduces
the longest measured autocorrelation time from 1623 to 27 sweeps in an
otherwise identical pilot.  Four independently seeded chains share a common
parallel-tempering ladder.

\subsection{Order and defect observables}
\label{sec:methods-observables}

The intensive specific heat plotted in Fig.~\ref{fig:temperature}(a) is
obtained from fluctuations of the extensive energy $E$ (with $k_B=1$),
\begin{equation}
 c_V=\frac{C_V}{N}
 =\frac{\av{E^2}-\av{E}^2}{N T^2}.
 \label{eq:specific-heat}
\end{equation}

For the order-parameter amplitudes in Fig.~\ref{fig:temperature}(b), we first
define the three component magnetizations and the scalar chirality
magnetization by
\begin{equation}
\begin{aligned}
 m_{s,a}&=\frac{1}{N}\sum_i s_{a,i},
 &m_s^2&=\frac{1}{3}\sum_{a=1}^{3}m_{s,a}^2,\\
 m_\tau&=\frac{1}{N}\sum_i\tau_i.
\end{aligned}\label{eq:sign-order-parameters}
\end{equation}
The plotted quantities are the thermal root-mean-square amplitudes
$\sqrt{\av{m_s^2}}$ and $\sqrt{\av{m_\tau^2}}$.  This normalization makes
both approach one in a uniform ordered domain, while the squares avoid
selecting a particular member of the symmetry-related finite-size ensemble.

For the sign and soft-spin models, we define the common structure-factor
notation
\begin{equation}
 S_X(\bk)=\frac{1}{n_XN}\sum_{\mu=1}^{n_X}
 \left\langle\left|\sum_i X_{\mu i}e^{i\bk\cdot\br_i}\right|^2\right\rangle,
 \label{eq:structure-factor}
\end{equation}
where $X\in\{s,\Phi\}$ denotes a three-component field with $n_X=3$, while
$X\in\{\tau,\chi\}$ denotes a scalar field with $n_X=1$ and
$X_{1i}\equiv X_i$.  Let $\bm b_1,\bm b_2$ be reciprocal primitive vectors,
$\bm a_\mu\cdot\bm b_\nu=2\pi\delta_{\mu\nu}$.  In the sign-model estimator,
$S_X(\bk_{\min})$ is shorthand for the average at $\bm b_1/L$ and
$\bm b_2/L$.  In the soft-spin estimator it is averaged over $\bm b_1/L$,
$\bm b_2/L$, and $(\bm b_1+\bm b_2)/L$; these are the shortest nonzero
reciprocal-lattice directions used in the respective measurements.

For the sign model we use the dimensionless correlation ratio
\begin{equation}
 R_X(L,T)=1-\frac{S_X(\bk_{\min})}{S_X(0)},
 \qquad X\in\{s,\tau\}.
 \label{eq:correlation-ratio}
\end{equation}
Crossings of $R_s$ and $R_\tau$ locate the two phase boundaries.  To expose
their defect content, we also count the number of reversed primary signs on a
nearest-neighbor bond,
\begin{equation}
\begin{aligned}
 h_{ij}&=\sum_{a=1}^{3}\frac{1-s_{a,i}s_{a,j}}{2},\\
 \rho_s&=\frac{1}{3N}\sum_{\langle ij\rangle}
 \delta_{h_{ij},2},\\
 \rho_\tau&=\frac{1}{3N}\sum_{\langle ij\rangle}
 \frac{1-\tau_i\tau_j}{2}.
\end{aligned}\label{eq:wall-densities}
\end{equation}
Thus $h_{ij}=2$ identifies a translation wall that preserves $\tau$, whereas
odd $h_{ij}$ reverses $\tau$.  Independent random signs give
$\rho_s=3/8$ and $\rho_\tau=1/2$.  These local quantities diagnose wall
proliferation but do not locate the transitions.  We additionally compute the
Binder ratios \cite{Maccari2023},
\begin{equation}
 U_s=\frac{3\av{m_s^4}}{5\av{m_s^2}^2},\qquad
 U_\tau=\frac{\av{m_\tau^4}}{3\av{m_\tau^2}^2}.
 \label{eq:sign-binder}
\end{equation}
The second expression is the scalar convention of Ref.~\cite{Maccari2023};
the factor $5/3$ in the first is the Gaussian fourth-moment ratio for a
three-component vector; here $m_s^4\equiv(m_s^2)^2$.  Accordingly,
$U_s\to3/5$ and $U_\tau\to1/3$ in their
rigid ordered limits, while both approach one in a Gaussian disordered phase.
Near the $q=2$ primary transition, $U_s(T)$ is nonmonotonic and has two
successive-size intersections.  We therefore use the unique $R_s$ crossing in
the main analysis and retain Binder results as Supplementary checks.  The
monotonic $U_\tau$ curves give a unique crossing consistent with $R_\tau$.

For the real-space diagnostic in Fig.~\ref{fig:correlations}, we measure
\begin{equation}
 C_s(r)=\av{s_{a,i}s_{a,i+r\hat e_d}}_{i,a,d},\quad
 C_\tau(r)=\av{\tau_i\tau_{i+r\hat e_d}}_{i,d},
 \label{eq:real-correlations}
\end{equation}
where $r$ is an integer separation and
$\hat e_d\in\{\bm a_1,\bm a_2,\bm a_1-\bm a_2\}$ runs over the three positive
triangular-lattice directions; the subscripts on $\av{\cdots}$ denote uniform
averages over the listed site, component, and direction indices.  Both
correlators equal one at $r=0$.  For the soft-spin model, the corresponding
dimensionless observable is the second-moment ratio
\begin{equation}
 \frac{\xi_X}{L}=\frac{1}{2L\sin(\pi/L)}
 \left[\frac{S_X(0)}{S_X(\bk_{\min})}-1\right]^{1/2}.
 \label{eq:xi-over-l}
\end{equation}
Here $X\in\{\Phi,\chi\}$ and $\xi_X$ is the second-moment correlation length
in channel $X$.
Successive-size crossings locate the transitions; Binder ratios,
susceptibilities, and structure-factor scaling test the resulting phase
assignment.  No boundary is inferred from a single-size peak.

\subsection{Convergence and statistical analysis}
\label{sec:methods-statistics}

Time series are blocked at a scale chosen to exceed the integrated autocorrelation 
time. Bootstrap replicas first resample independent chains and then whole blocks 
within each selected chain. For parallel-tempering data, each block contains the 
synchronized measurements from the complete temperature ladder; the same resampled 
block indices are therefore applied to all temperature slots and primitive observables 
before nonlinear ratios and crossing temperatures are reconstructed. Fixed-temperature 
scans are instead resampled independently at each temperature. Different lattice sizes 
are resampled independently, while the bootstrap curve for a given size is reused in 
adjacent size-pair crossings.  We monitor split-$\hat R$
\cite{Vehtari2021}, effective sample size, reliable block count, and replica
round trips as complementary convergence tests.  Near a transport-limited
sign-model crossing, the uncertainty is floored at half the native
temperature-grid spacing.  Ferrenberg--Swendsen/WHAM reweighting of energy
histograms, where WHAM denotes the weighted histogram analysis method, tests
for phase coexistence \cite{FerrenbergSwendsen1989}.

\subsection{Use of artificial intelligence tools}

OpenAI Codex with GPT-5.6 Sol was used interactively to assist with software
implementation and debugging, data-analysis workflows, and plotting scripts.
The authors specified the physical models, algorithms, and numerical protocols
and verified the AI-assisted output through manual code review, automated unit
and regression tests, cross-comparisons between independent CPU and Metal/GPU
implementations, and equilibrium comparisons between embedded Swendsen--Wang
cluster and independent single-site heat-bath samplers.  The simulations were
additionally benchmarked against the exactly known triangular-lattice Ising
and four-state Potts limiting cases.  All simulations were executed and all
reported numerical results and figures were inspected by the authors.

\bibliography{references}

\begin{thebibliography}{46}%
\makeatletter
\providecommand \@ifxundefined [1]{%
 \@ifx{#1\undefined}
}%
\providecommand \@ifnum [1]{%
 \ifnum #1\expandafter \@firstoftwo
 \else \expandafter \@secondoftwo
 \fi
}%
\providecommand \@ifx [1]{%
 \ifx #1\expandafter \@firstoftwo
 \else \expandafter \@secondoftwo
 \fi
}%
\providecommand \natexlab [1]{#1}%
\providecommand \enquote  [1]{``#1''}%
\providecommand \bibnamefont  [1]{#1}%
\providecommand \bibfnamefont [1]{#1}%
\providecommand \citenamefont [1]{#1}%
\providecommand \href@noop [0]{\@secondoftwo}%
\providecommand \href [0]{\begingroup \@sanitize@url \@href}%
\providecommand \@href[1]{\@@startlink{#1}\@@href}%
\providecommand \@@href[1]{\endgroup#1\@@endlink}%
\providecommand \@sanitize@url [0]{\catcode `\\12\catcode `\$12\catcode `\&12\catcode `\#12\catcode `\^12\catcode `\_12\catcode `\%12\relax}%
\providecommand \@@startlink[1]{}%
\providecommand \@@endlink[0]{}%
\providecommand \url  [0]{\begingroup\@sanitize@url \@url }%
\providecommand \@url [1]{\endgroup\@href {#1}{\urlprefix }}%
\providecommand \urlprefix  [0]{URL }%
\providecommand \Eprint [0]{\href }%
\providecommand \doibase [0]{https://doi.org/}%
\providecommand \selectlanguage [0]{\@gobble}%
\providecommand \bibinfo  [0]{\@secondoftwo}%
\providecommand \bibfield  [0]{\@secondoftwo}%
\providecommand \translation [1]{[#1]}%
\providecommand \BibitemOpen [0]{}%
\providecommand \bibitemStop [0]{}%
\providecommand \bibitemNoStop [0]{.\EOS\space}%
\providecommand \EOS [0]{\spacefactor3000\relax}%
\providecommand \BibitemShut  [1]{\csname bibitem#1\endcsname}%
\let\auto@bib@innerbib\@empty
\bibitem [{\citenamefont {Ortiz}\ \emph {et~al.}(2019)\citenamefont {Ortiz}, \citenamefont {Gomes}, \citenamefont {Morey}, \citenamefont {Winiarski}, \citenamefont {Bordelon}, \citenamefont {Mangum}, \citenamefont {Oswald}, \citenamefont {Rodriguez-Rivera}, \citenamefont {Neilson}, \citenamefont {Wilson}, \citenamefont {Ertekin}, \citenamefont {McQueen},\ and\ \citenamefont {Toberer}}]{Ortiz2019}%
  \BibitemOpen
  \bibfield  {author} {\bibinfo {author} {\bibfnamefont {B.~R.}\ \bibnamefont {Ortiz}}, \bibinfo {author} {\bibfnamefont {L.~C.}\ \bibnamefont {Gomes}}, \bibinfo {author} {\bibfnamefont {J.~R.}\ \bibnamefont {Morey}}, \bibinfo {author} {\bibfnamefont {M.}~\bibnamefont {Winiarski}}, \bibinfo {author} {\bibfnamefont {M.}~\bibnamefont {Bordelon}}, \bibinfo {author} {\bibfnamefont {J.~S.}\ \bibnamefont {Mangum}}, \bibinfo {author} {\bibfnamefont {I.~W.~H.}\ \bibnamefont {Oswald}}, \bibinfo {author} {\bibfnamefont {J.~A.}\ \bibnamefont {Rodriguez-Rivera}}, \bibinfo {author} {\bibfnamefont {J.~R.}\ \bibnamefont {Neilson}}, \bibinfo {author} {\bibfnamefont {S.~D.}\ \bibnamefont {Wilson}}, \bibinfo {author} {\bibfnamefont {E.}~\bibnamefont {Ertekin}}, \bibinfo {author} {\bibfnamefont {T.~M.}\ \bibnamefont {McQueen}},\ and\ \bibinfo {author} {\bibfnamefont {E.~S.}\ \bibnamefont {Toberer}},\ }\bibfield  {title} {\bibinfo {title} {New kagome prototype materials: Discovery of {KV$_3$Sb$_5$}, {RbV$_3$Sb$_5$}, and {CsV$_3$Sb$_5$}},\ }\href {https://doi.org/10.1103/PhysRevMaterials.3.094407} {\bibfield  {journal} {\bibinfo  {journal} {Physical Review Materials}\ }\textbf {\bibinfo {volume} {3}},\ \bibinfo {pages} {094407} (\bibinfo {year} {2019})}\BibitemShut {NoStop}%
\bibitem [{\citenamefont {Ortiz}\ \emph {et~al.}(2020)\citenamefont {Ortiz}, \citenamefont {Teicher}, \citenamefont {Hu}, \citenamefont {Zuo}, \citenamefont {Sarte}, \citenamefont {Schueller}, \citenamefont {Abeykoon}, \citenamefont {Krogstad}, \citenamefont {Rosenkranz}, \citenamefont {Osborn}, \citenamefont {Seshadri}, \citenamefont {Balents}, \citenamefont {He},\ and\ \citenamefont {Wilson}}]{Ortiz2020}%
  \BibitemOpen
  \bibfield  {author} {\bibinfo {author} {\bibfnamefont {B.~R.}\ \bibnamefont {Ortiz}}, \bibinfo {author} {\bibfnamefont {S.~M.~L.}\ \bibnamefont {Teicher}}, \bibinfo {author} {\bibfnamefont {Y.}~\bibnamefont {Hu}}, \bibinfo {author} {\bibfnamefont {J.~L.}\ \bibnamefont {Zuo}}, \bibinfo {author} {\bibfnamefont {P.~M.}\ \bibnamefont {Sarte}}, \bibinfo {author} {\bibfnamefont {E.~C.}\ \bibnamefont {Schueller}}, \bibinfo {author} {\bibfnamefont {A.~M.~M.}\ \bibnamefont {Abeykoon}}, \bibinfo {author} {\bibfnamefont {M.~J.}\ \bibnamefont {Krogstad}}, \bibinfo {author} {\bibfnamefont {S.}~\bibnamefont {Rosenkranz}}, \bibinfo {author} {\bibfnamefont {R.}~\bibnamefont {Osborn}}, \bibinfo {author} {\bibfnamefont {R.}~\bibnamefont {Seshadri}}, \bibinfo {author} {\bibfnamefont {L.}~\bibnamefont {Balents}}, \bibinfo {author} {\bibfnamefont {J.}~\bibnamefont {He}},\ and\ \bibinfo {author} {\bibfnamefont {S.~D.}\ \bibnamefont {Wilson}},\ }\bibfield  {title} {\bibinfo {title} {{CsV$_3$Sb$_5$}: A {$\mathbb{Z}_2$} topological kagome metal with a superconducting ground state},\ }\href {https://doi.org/10.1103/PhysRevLett.125.247002} {\bibfield  {journal} {\bibinfo  {journal} {Physical Review Letters}\ }\textbf {\bibinfo {volume} {125}},\ \bibinfo {pages} {247002} (\bibinfo {year} {2020})}\BibitemShut {NoStop}%
\bibitem [{\citenamefont {Neupert}\ \emph {et~al.}(2022)\citenamefont {Neupert}, \citenamefont {Denner}, \citenamefont {Yin}, \citenamefont {Thomale},\ and\ \citenamefont {Hasan}}]{Neupert2022}%
  \BibitemOpen
  \bibfield  {author} {\bibinfo {author} {\bibfnamefont {T.}~\bibnamefont {Neupert}}, \bibinfo {author} {\bibfnamefont {M.~M.}\ \bibnamefont {Denner}}, \bibinfo {author} {\bibfnamefont {J.-X.}\ \bibnamefont {Yin}}, \bibinfo {author} {\bibfnamefont {R.}~\bibnamefont {Thomale}},\ and\ \bibinfo {author} {\bibfnamefont {M.~Z.}\ \bibnamefont {Hasan}},\ }\bibfield  {title} {\bibinfo {title} {Charge order and superconductivity in kagome materials},\ }\href {https://doi.org/10.1038/s41567-021-01404-y} {\bibfield  {journal} {\bibinfo  {journal} {Nature Physics}\ }\textbf {\bibinfo {volume} {18}},\ \bibinfo {pages} {137} (\bibinfo {year} {2022})}\BibitemShut {NoStop}%
\bibitem [{\citenamefont {Zhao}\ \emph {et~al.}(2021)\citenamefont {Zhao}, \citenamefont {Li}, \citenamefont {Ortiz}, \citenamefont {Teicher}, \citenamefont {Park}, \citenamefont {Ye}, \citenamefont {Wang}, \citenamefont {Balents}, \citenamefont {Wilson},\ and\ \citenamefont {Zeljkovic}}]{Zhao2021}%
  \BibitemOpen
  \bibfield  {author} {\bibinfo {author} {\bibfnamefont {H.}~\bibnamefont {Zhao}}, \bibinfo {author} {\bibfnamefont {H.}~\bibnamefont {Li}}, \bibinfo {author} {\bibfnamefont {B.~R.}\ \bibnamefont {Ortiz}}, \bibinfo {author} {\bibfnamefont {S.~M.~L.}\ \bibnamefont {Teicher}}, \bibinfo {author} {\bibfnamefont {T.}~\bibnamefont {Park}}, \bibinfo {author} {\bibfnamefont {M.}~\bibnamefont {Ye}}, \bibinfo {author} {\bibfnamefont {Z.}~\bibnamefont {Wang}}, \bibinfo {author} {\bibfnamefont {L.}~\bibnamefont {Balents}}, \bibinfo {author} {\bibfnamefont {S.~D.}\ \bibnamefont {Wilson}},\ and\ \bibinfo {author} {\bibfnamefont {I.}~\bibnamefont {Zeljkovic}},\ }\bibfield  {title} {\bibinfo {title} {Cascade of correlated electron states in the kagome superconductor {CsV$_3$Sb$_5$}},\ }\href {https://doi.org/10.1038/s41586-021-03946-w} {\bibfield  {journal} {\bibinfo  {journal} {Nature}\ }\textbf {\bibinfo {volume} {599}},\ \bibinfo {pages} {216} (\bibinfo {year} {2021})}\BibitemShut {NoStop}%
\bibitem [{\citenamefont {Jiang}\ \emph {et~al.}(2021)\citenamefont {Jiang}, \citenamefont {Yin}, \citenamefont {Denner}, \citenamefont {Shumiya}, \citenamefont {Ortiz}, \citenamefont {Xu}, \citenamefont {Guguchia}, \citenamefont {He}, \citenamefont {Hossain}, \citenamefont {Liu}, \citenamefont {Ruff}, \citenamefont {Kautzsch}, \citenamefont {Zhang}, \citenamefont {Chang}, \citenamefont {Belopolski}, \citenamefont {Zhang}, \citenamefont {Cochran}, \citenamefont {Multer}, \citenamefont {Litskevich}, \citenamefont {Cheng}, \citenamefont {Yang}, \citenamefont {Wang}, \citenamefont {Thomale}, \citenamefont {Neupert}, \citenamefont {Wilson},\ and\ \citenamefont {Hasan}}]{Jiang2021}%
  \BibitemOpen
  \bibfield  {author} {\bibinfo {author} {\bibfnamefont {Y.-X.}\ \bibnamefont {Jiang}}, \bibinfo {author} {\bibfnamefont {J.-X.}\ \bibnamefont {Yin}}, \bibinfo {author} {\bibfnamefont {M.~M.}\ \bibnamefont {Denner}}, \bibinfo {author} {\bibfnamefont {N.}~\bibnamefont {Shumiya}}, \bibinfo {author} {\bibfnamefont {B.~R.}\ \bibnamefont {Ortiz}}, \bibinfo {author} {\bibfnamefont {G.}~\bibnamefont {Xu}}, \bibinfo {author} {\bibfnamefont {Z.}~\bibnamefont {Guguchia}}, \bibinfo {author} {\bibfnamefont {J.}~\bibnamefont {He}}, \bibinfo {author} {\bibfnamefont {M.~S.}\ \bibnamefont {Hossain}}, \bibinfo {author} {\bibfnamefont {X.}~\bibnamefont {Liu}}, \bibinfo {author} {\bibfnamefont {J.}~\bibnamefont {Ruff}}, \bibinfo {author} {\bibfnamefont {L.}~\bibnamefont {Kautzsch}}, \bibinfo {author} {\bibfnamefont {S.~S.}\ \bibnamefont {Zhang}}, \bibinfo {author} {\bibfnamefont {G.}~\bibnamefont {Chang}}, \bibinfo {author} {\bibfnamefont {I.}~\bibnamefont {Belopolski}}, \bibinfo {author} {\bibfnamefont {Q.}~\bibnamefont {Zhang}}, \bibinfo {author} {\bibfnamefont {T.~A.}\ \bibnamefont {Cochran}}, \bibinfo {author} {\bibfnamefont {D.}~\bibnamefont {Multer}}, \bibinfo {author} {\bibfnamefont {M.}~\bibnamefont {Litskevich}}, \bibinfo {author} {\bibfnamefont {Z.-J.}\ \bibnamefont {Cheng}}, \bibinfo {author} {\bibfnamefont {X.~P.}\ \bibnamefont {Yang}}, \bibinfo {author} {\bibfnamefont {Z.}~\bibnamefont {Wang}}, \bibinfo {author} {\bibfnamefont {R.}~\bibnamefont {Thomale}}, \bibinfo {author} {\bibfnamefont {T.}~\bibnamefont {Neupert}}, \bibinfo {author} {\bibfnamefont {S.~D.}\ \bibnamefont {Wilson}},\ and\ \bibinfo {author} {\bibfnamefont {M.~Z.}\ \bibnamefont {Hasan}},\ }\bibfield  {title} {\bibinfo {title} {Unconventional chiral charge order in kagome superconductor {KV$_3$Sb$_5$}},\ }\href {https://doi.org/10.1038/s41563-021-01034-y} {\bibfield  {journal} {\bibinfo  {journal} {Nature Materials}\ }\textbf {\bibinfo {volume} {20}},\ \bibinfo {pages} {1353} (\bibinfo {year} {2021})}\BibitemShut {NoStop}%
\bibitem [{\citenamefont {Mielke}\ \emph {et~al.}(2022)\citenamefont {Mielke}, \citenamefont {Das}, \citenamefont {Yin}, \citenamefont {Liu}, \citenamefont {Gupta}, \citenamefont {Jiang}, \citenamefont {Medarde}, \citenamefont {Wu}, \citenamefont {Lei}, \citenamefont {Chang}, \citenamefont {Dai}, \citenamefont {Si}, \citenamefont {Miao}, \citenamefont {Thomale}, \citenamefont {Neupert}, \citenamefont {Shi}, \citenamefont {Khasanov}, \citenamefont {Hasan}, \citenamefont {Luetkens},\ and\ \citenamefont {Guguchia}}]{Mielke2022}%
  \BibitemOpen
  \bibfield  {author} {\bibinfo {author} {\bibfnamefont {C.}~\bibnamefont {Mielke}}, \bibinfo {author} {\bibfnamefont {D.}~\bibnamefont {Das}}, \bibinfo {author} {\bibfnamefont {J.-X.}\ \bibnamefont {Yin}}, \bibinfo {author} {\bibfnamefont {H.}~\bibnamefont {Liu}}, \bibinfo {author} {\bibfnamefont {R.}~\bibnamefont {Gupta}}, \bibinfo {author} {\bibfnamefont {Y.-X.}\ \bibnamefont {Jiang}}, \bibinfo {author} {\bibfnamefont {M.}~\bibnamefont {Medarde}}, \bibinfo {author} {\bibfnamefont {X.}~\bibnamefont {Wu}}, \bibinfo {author} {\bibfnamefont {H.~C.}\ \bibnamefont {Lei}}, \bibinfo {author} {\bibfnamefont {J.}~\bibnamefont {Chang}}, \bibinfo {author} {\bibfnamefont {P.}~\bibnamefont {Dai}}, \bibinfo {author} {\bibfnamefont {Q.}~\bibnamefont {Si}}, \bibinfo {author} {\bibfnamefont {H.}~\bibnamefont {Miao}}, \bibinfo {author} {\bibfnamefont {R.}~\bibnamefont {Thomale}}, \bibinfo {author} {\bibfnamefont {T.}~\bibnamefont {Neupert}}, \bibinfo {author} {\bibfnamefont {Y.}~\bibnamefont {Shi}}, \bibinfo {author} {\bibfnamefont {R.}~\bibnamefont {Khasanov}}, \bibinfo {author} {\bibfnamefont {M.~Z.}\ \bibnamefont {Hasan}}, \bibinfo {author} {\bibfnamefont {H.}~\bibnamefont {Luetkens}},\ and\ \bibinfo {author} {\bibfnamefont {Z.}~\bibnamefont {Guguchia}},\ }\bibfield  {title} {\bibinfo {title} {Time-reversal symmetry-breaking charge order in a kagome superconductor},\ }\href {https://doi.org/10.1038/s41586-021-04327-z} {\bibfield  {journal} {\bibinfo  {journal} {Nature}\ }\textbf {\bibinfo {volume} {602}},\ \bibinfo {pages} {245} (\bibinfo {year} {2022})}\BibitemShut {NoStop}%
\bibitem [{\citenamefont {Chen}\ \emph {et~al.}(2022)\citenamefont {Chen}, \citenamefont {Chen}, \citenamefont {Schnelle}, \citenamefont {Felser},\ and\ \citenamefont {Gaulin}}]{Chen2022}%
  \BibitemOpen
  \bibfield  {author} {\bibinfo {author} {\bibfnamefont {Q.}~\bibnamefont {Chen}}, \bibinfo {author} {\bibfnamefont {D.}~\bibnamefont {Chen}}, \bibinfo {author} {\bibfnamefont {W.}~\bibnamefont {Schnelle}}, \bibinfo {author} {\bibfnamefont {C.}~\bibnamefont {Felser}},\ and\ \bibinfo {author} {\bibfnamefont {B.~D.}\ \bibnamefont {Gaulin}},\ }\bibfield  {title} {\bibinfo {title} {Charge density wave order and fluctuations above {$T_{\mathrm{CDW}}$} and below superconducting {$T_c$} in the kagome metal {CsV$_3$Sb$_5$}},\ }\href {https://doi.org/10.1103/PhysRevLett.129.056401} {\bibfield  {journal} {\bibinfo  {journal} {Physical Review Letters}\ }\textbf {\bibinfo {volume} {129}},\ \bibinfo {pages} {056401} (\bibinfo {year} {2022})}\BibitemShut {NoStop}%
\bibitem [{\citenamefont {Subires}\ \emph {et~al.}(2023)\citenamefont {Subires}, \citenamefont {Korshunov}, \citenamefont {Said}, \citenamefont {S\'anchez}, \citenamefont {Ortiz}, \citenamefont {Wilson}, \citenamefont {Bosak},\ and\ \citenamefont {Blanco-Canosa}}]{Subires2023}%
  \BibitemOpen
  \bibfield  {author} {\bibinfo {author} {\bibfnamefont {D.}~\bibnamefont {Subires}}, \bibinfo {author} {\bibfnamefont {A.}~\bibnamefont {Korshunov}}, \bibinfo {author} {\bibfnamefont {A.~H.}\ \bibnamefont {Said}}, \bibinfo {author} {\bibfnamefont {L.}~\bibnamefont {S\'anchez}}, \bibinfo {author} {\bibfnamefont {B.~R.}\ \bibnamefont {Ortiz}}, \bibinfo {author} {\bibfnamefont {S.~D.}\ \bibnamefont {Wilson}}, \bibinfo {author} {\bibfnamefont {A.}~\bibnamefont {Bosak}},\ and\ \bibinfo {author} {\bibfnamefont {S.}~\bibnamefont {Blanco-Canosa}},\ }\bibfield  {title} {\bibinfo {title} {Order-disorder charge density wave instability in the kagome metal {{(Cs,Rb)V$_3$Sb$_5$}}},\ }\href {https://doi.org/10.1038/s41467-023-36668-w} {\bibfield  {journal} {\bibinfo  {journal} {Nature Communications}\ }\textbf {\bibinfo {volume} {14}},\ \bibinfo {pages} {1015} (\bibinfo {year} {2023})}\BibitemShut {NoStop}%
\bibitem [{\citenamefont {Graham}\ \emph {et~al.}(2024)\citenamefont {Graham}, \citenamefont {{Mielke III}}, \citenamefont {Das}, \citenamefont {Morresi}, \citenamefont {Sazgari}, \citenamefont {Suter}, \citenamefont {Prokscha}, \citenamefont {Deng}, \citenamefont {Khasanov}, \citenamefont {Wilson}, \citenamefont {Salinas}, \citenamefont {Martins}, \citenamefont {Zhong}, \citenamefont {Okazaki}, \citenamefont {Wang}, \citenamefont {Hasan}, \citenamefont {Fischer}, \citenamefont {Neupert}, \citenamefont {Yin}, \citenamefont {Sanna}, \citenamefont {Luetkens}, \citenamefont {Salman}, \citenamefont {Bonf\`a},\ and\ \citenamefont {Guguchia}}]{Graham2024}%
  \BibitemOpen
  \bibfield  {author} {\bibinfo {author} {\bibfnamefont {J.~N.}\ \bibnamefont {Graham}}, \bibinfo {author} {\bibfnamefont {C.}~\bibnamefont {{Mielke III}}}, \bibinfo {author} {\bibfnamefont {D.}~\bibnamefont {Das}}, \bibinfo {author} {\bibfnamefont {T.}~\bibnamefont {Morresi}}, \bibinfo {author} {\bibfnamefont {V.}~\bibnamefont {Sazgari}}, \bibinfo {author} {\bibfnamefont {A.}~\bibnamefont {Suter}}, \bibinfo {author} {\bibfnamefont {T.}~\bibnamefont {Prokscha}}, \bibinfo {author} {\bibfnamefont {H.}~\bibnamefont {Deng}}, \bibinfo {author} {\bibfnamefont {R.}~\bibnamefont {Khasanov}}, \bibinfo {author} {\bibfnamefont {S.~D.}\ \bibnamefont {Wilson}}, \bibinfo {author} {\bibfnamefont {A.~C.}\ \bibnamefont {Salinas}}, \bibinfo {author} {\bibfnamefont {M.~M.}\ \bibnamefont {Martins}}, \bibinfo {author} {\bibfnamefont {Y.}~\bibnamefont {Zhong}}, \bibinfo {author} {\bibfnamefont {K.}~\bibnamefont {Okazaki}}, \bibinfo {author} {\bibfnamefont {Z.}~\bibnamefont {Wang}}, \bibinfo {author} {\bibfnamefont {M.~Z.}\ \bibnamefont {Hasan}}, \bibinfo {author} {\bibfnamefont {M.~H.}\ \bibnamefont {Fischer}}, \bibinfo {author} {\bibfnamefont {T.}~\bibnamefont {Neupert}}, \bibinfo {author} {\bibfnamefont {J.-X.}\ \bibnamefont {Yin}}, \bibinfo {author} {\bibfnamefont {S.}~\bibnamefont {Sanna}}, \bibinfo {author} {\bibfnamefont {H.}~\bibnamefont {Luetkens}}, \bibinfo {author} {\bibfnamefont {Z.}~\bibnamefont {Salman}}, \bibinfo {author} {\bibfnamefont {P.}~\bibnamefont {Bonf\`a}},\ and\ \bibinfo {author} {\bibfnamefont {Z.}~\bibnamefont {Guguchia}},\ }\bibfield  {title} {\bibinfo {title} {Depth-dependent study of time-reversal symmetry-breaking in the kagome superconductor {{$A$V$_3$Sb$_5$}}},\ }\href {https://doi.org/10.1038/s41467-024-52688-6} {\bibfield  {journal} {\bibinfo  {journal} {Nature Communications}\ }\textbf {\bibinfo {volume} {15}},\ \bibinfo {pages} {8978} (\bibinfo {year} {2024})}\BibitemShut {NoStop}%
\bibitem [{\citenamefont {Denner}\ \emph {et~al.}(2021)\citenamefont {Denner}, \citenamefont {Thomale},\ and\ \citenamefont {Neupert}}]{Denner2021}%
  \BibitemOpen
  \bibfield  {author} {\bibinfo {author} {\bibfnamefont {M.~M.}\ \bibnamefont {Denner}}, \bibinfo {author} {\bibfnamefont {R.}~\bibnamefont {Thomale}},\ and\ \bibinfo {author} {\bibfnamefont {T.}~\bibnamefont {Neupert}},\ }\bibfield  {title} {\bibinfo {title} {Analysis of charge order in the kagome metal {$A$V$_3$Sb$_5$} {($A={}$K, Rb, Cs)}},\ }\href {https://doi.org/10.1103/PhysRevLett.127.217601} {\bibfield  {journal} {\bibinfo  {journal} {Physical Review Letters}\ }\textbf {\bibinfo {volume} {127}},\ \bibinfo {pages} {217601} (\bibinfo {year} {2021})}\BibitemShut {NoStop}%
\bibitem [{\citenamefont {Park}\ \emph {et~al.}(2021)\citenamefont {Park}, \citenamefont {Ye},\ and\ \citenamefont {Balents}}]{Park2021}%
  \BibitemOpen
  \bibfield  {author} {\bibinfo {author} {\bibfnamefont {T.}~\bibnamefont {Park}}, \bibinfo {author} {\bibfnamefont {M.}~\bibnamefont {Ye}},\ and\ \bibinfo {author} {\bibfnamefont {L.}~\bibnamefont {Balents}},\ }\bibfield  {title} {\bibinfo {title} {Electronic instabilities of kagome metals: Saddle points and landau theory},\ }\href {https://doi.org/10.1103/PhysRevB.104.035142} {\bibfield  {journal} {\bibinfo  {journal} {Physical Review B}\ }\textbf {\bibinfo {volume} {104}},\ \bibinfo {pages} {035142} (\bibinfo {year} {2021})}\BibitemShut {NoStop}%
\bibitem [{\citenamefont {Lin}\ and\ \citenamefont {Nandkishore}(2021)}]{LinNandkishore2021}%
  \BibitemOpen
  \bibfield  {author} {\bibinfo {author} {\bibfnamefont {Y.-P.}\ \bibnamefont {Lin}}\ and\ \bibinfo {author} {\bibfnamefont {R.~M.}\ \bibnamefont {Nandkishore}},\ }\bibfield  {title} {\bibinfo {title} {Complex charge density waves at van hove singularity on hexagonal lattices: Haldane-model phase diagram and potential realization in the kagome metals {$A$V$_3$Sb$_5$} {($A={}$K, Rb, Cs)}},\ }\href {https://doi.org/10.1103/PhysRevB.104.045122} {\bibfield  {journal} {\bibinfo  {journal} {Physical Review B}\ }\textbf {\bibinfo {volume} {104}},\ \bibinfo {pages} {045122} (\bibinfo {year} {2021})}\BibitemShut {NoStop}%
\bibitem [{\citenamefont {Li}\ \emph {et~al.}(2024)\citenamefont {Li}, \citenamefont {Kim},\ and\ \citenamefont {Kee}}]{Li2024}%
  \BibitemOpen
  \bibfield  {author} {\bibinfo {author} {\bibfnamefont {H.}~\bibnamefont {Li}}, \bibinfo {author} {\bibfnamefont {Y.~B.}\ \bibnamefont {Kim}},\ and\ \bibinfo {author} {\bibfnamefont {H.-Y.}\ \bibnamefont {Kee}},\ }\bibfield  {title} {\bibinfo {title} {Intertwined van hove singularities as a mechanism for loop current order in kagome metals},\ }\href {https://doi.org/10.1103/PhysRevLett.132.146501} {\bibfield  {journal} {\bibinfo  {journal} {Physical Review Letters}\ }\textbf {\bibinfo {volume} {132}},\ \bibinfo {pages} {146501} (\bibinfo {year} {2024})}\BibitemShut {NoStop}%
\bibitem [{\citenamefont {Zhan}\ \emph {et~al.}(2026)\citenamefont {Zhan}, \citenamefont {Hohmann}, \citenamefont {D\"urrnagel}, \citenamefont {Fu}, \citenamefont {Zhou}, \citenamefont {Wang}, \citenamefont {Thomale}, \citenamefont {Wu},\ and\ \citenamefont {Hu}}]{Zhan2026}%
  \BibitemOpen
  \bibfield  {author} {\bibinfo {author} {\bibfnamefont {J.}~\bibnamefont {Zhan}}, \bibinfo {author} {\bibfnamefont {H.}~\bibnamefont {Hohmann}}, \bibinfo {author} {\bibfnamefont {M.}~\bibnamefont {D\"urrnagel}}, \bibinfo {author} {\bibfnamefont {R.}~\bibnamefont {Fu}}, \bibinfo {author} {\bibfnamefont {S.}~\bibnamefont {Zhou}}, \bibinfo {author} {\bibfnamefont {Z.}~\bibnamefont {Wang}}, \bibinfo {author} {\bibfnamefont {R.}~\bibnamefont {Thomale}}, \bibinfo {author} {\bibfnamefont {X.}~\bibnamefont {Wu}},\ and\ \bibinfo {author} {\bibfnamefont {J.}~\bibnamefont {Hu}},\ }\bibfield  {title} {\bibinfo {title} {Loop current order on the kagome lattice},\ }\href {https://doi.org/10.1103/5vyy-rj6v} {\bibfield  {journal} {\bibinfo  {journal} {Physical Review Letters}\ }\textbf {\bibinfo {volume} {136}},\ \bibinfo {pages} {126001} (\bibinfo {year} {2026})}\BibitemShut {NoStop}%
\bibitem [{\citenamefont {Feng}\ \emph {et~al.}(2021)\citenamefont {Feng}, \citenamefont {Zhang}, \citenamefont {Jiang},\ and\ \citenamefont {Hu}}]{Feng2021}%
  \BibitemOpen
  \bibfield  {author} {\bibinfo {author} {\bibfnamefont {X.}~\bibnamefont {Feng}}, \bibinfo {author} {\bibfnamefont {Y.}~\bibnamefont {Zhang}}, \bibinfo {author} {\bibfnamefont {K.}~\bibnamefont {Jiang}},\ and\ \bibinfo {author} {\bibfnamefont {J.}~\bibnamefont {Hu}},\ }\bibfield  {title} {\bibinfo {title} {Low-energy effective theory and symmetry classification of flux phases on the kagome lattice},\ }\href {https://doi.org/10.1103/PhysRevB.104.165136} {\bibfield  {journal} {\bibinfo  {journal} {Physical Review B}\ }\textbf {\bibinfo {volume} {104}},\ \bibinfo {pages} {165136} (\bibinfo {year} {2021})}\BibitemShut {NoStop}%
\bibitem [{\citenamefont {Christensen}\ \emph {et~al.}(2022)\citenamefont {Christensen}, \citenamefont {Birol}, \citenamefont {Andersen},\ and\ \citenamefont {Fernandes}}]{Christensen2022}%
  \BibitemOpen
  \bibfield  {author} {\bibinfo {author} {\bibfnamefont {M.~H.}\ \bibnamefont {Christensen}}, \bibinfo {author} {\bibfnamefont {T.}~\bibnamefont {Birol}}, \bibinfo {author} {\bibfnamefont {B.~M.}\ \bibnamefont {Andersen}},\ and\ \bibinfo {author} {\bibfnamefont {R.~M.}\ \bibnamefont {Fernandes}},\ }\bibfield  {title} {\bibinfo {title} {Loop currents in {$A$V$_3$Sb$_5$} kagome metals: Multipolar and toroidal magnetic orders},\ }\href {https://doi.org/10.1103/PhysRevB.106.144504} {\bibfield  {journal} {\bibinfo  {journal} {Physical Review B}\ }\textbf {\bibinfo {volume} {106}},\ \bibinfo {pages} {144504} (\bibinfo {year} {2022})}\BibitemShut {NoStop}%
\bibitem [{\citenamefont {Tazai}\ \emph {et~al.}(2023)\citenamefont {Tazai}, \citenamefont {Yamakawa},\ and\ \citenamefont {Kontani}}]{Tazai2023}%
  \BibitemOpen
  \bibfield  {author} {\bibinfo {author} {\bibfnamefont {R.}~\bibnamefont {Tazai}}, \bibinfo {author} {\bibfnamefont {Y.}~\bibnamefont {Yamakawa}},\ and\ \bibinfo {author} {\bibfnamefont {H.}~\bibnamefont {Kontani}},\ }\bibfield  {title} {\bibinfo {title} {Charge-loop current order and {$Z_3$} nematicity mediated by bond-order fluctuations in kagome metals},\ }\href {https://doi.org/10.1038/s41467-023-42952-6} {\bibfield  {journal} {\bibinfo  {journal} {Nature Communications}\ }\textbf {\bibinfo {volume} {14}},\ \bibinfo {pages} {7845} (\bibinfo {year} {2023})}\BibitemShut {NoStop}%
\bibitem [{\citenamefont {Cha}\ \emph {et~al.}(2026)\citenamefont {Cha}, \citenamefont {Lee}, \citenamefont {Sim}, \citenamefont {Sur}, \citenamefont {Kim}, \citenamefont {Han}, \citenamefont {Kim}, \citenamefont {Lee}, \citenamefont {Hyun}, \citenamefont {Lim}, \citenamefont {Ahn}, \citenamefont {Gim}, \citenamefont {Denlinger}, \citenamefont {Kim}, \citenamefont {Kim}, \citenamefont {Lee}, \citenamefont {Han},\ and\ \citenamefont {Kim}}]{Cha2026}%
  \BibitemOpen
  \bibfield  {author} {\bibinfo {author} {\bibfnamefont {J.}~\bibnamefont {Cha}}, \bibinfo {author} {\bibfnamefont {H.}~\bibnamefont {Lee}}, \bibinfo {author} {\bibfnamefont {S.}~\bibnamefont {Sim}}, \bibinfo {author} {\bibfnamefont {Y.}~\bibnamefont {Sur}}, \bibinfo {author} {\bibfnamefont {K.-T.}\ \bibnamefont {Kim}}, \bibinfo {author} {\bibfnamefont {J.-H.}\ \bibnamefont {Han}}, \bibinfo {author} {\bibfnamefont {S.-W.}\ \bibnamefont {Kim}}, \bibinfo {author} {\bibfnamefont {G.}~\bibnamefont {Lee}}, \bibinfo {author} {\bibfnamefont {J.}~\bibnamefont {Hyun}}, \bibinfo {author} {\bibfnamefont {C.-y.}\ \bibnamefont {Lim}}, \bibinfo {author} {\bibfnamefont {Y.}~\bibnamefont {Ahn}}, \bibinfo {author} {\bibfnamefont {S.}~\bibnamefont {Gim}}, \bibinfo {author} {\bibfnamefont {J.~D.}\ \bibnamefont {Denlinger}}, \bibinfo {author} {\bibfnamefont {S.}~\bibnamefont {Kim}}, \bibinfo {author} {\bibfnamefont {K.~H.}\ \bibnamefont {Kim}}, \bibinfo {author} {\bibfnamefont {S.}~\bibnamefont {Lee}}, \bibinfo {author} {\bibfnamefont {M.~J.}\ \bibnamefont {Han}},\ and\ \bibinfo {author} {\bibfnamefont {Y.}~\bibnamefont {Kim}},\ }\bibfield  {title} {\bibinfo {title} {Evidence of time-reversal symmetry breaking above the charge density wave order in a kagome metal},\ }\href {https://doi.org/10.1038/s41567-026-03331-2} {\bibfield  {journal} {\bibinfo  {journal} {Nature Physics}\ }\textbf {\bibinfo {volume} {22}},\ \bibinfo {pages} {1245} (\bibinfo {year} {2026})}\BibitemShut {NoStop}%
\bibitem [{\citenamefont {Tan}\ \emph {et~al.}(2021)\citenamefont {Tan}, \citenamefont {Liu}, \citenamefont {Wang},\ and\ \citenamefont {Yan}}]{Tan2021}%
  \BibitemOpen
  \bibfield  {author} {\bibinfo {author} {\bibfnamefont {H.}~\bibnamefont {Tan}}, \bibinfo {author} {\bibfnamefont {Y.}~\bibnamefont {Liu}}, \bibinfo {author} {\bibfnamefont {Z.}~\bibnamefont {Wang}},\ and\ \bibinfo {author} {\bibfnamefont {B.}~\bibnamefont {Yan}},\ }\bibfield  {title} {\bibinfo {title} {Charge density waves and electronic properties of superconducting kagome metals},\ }\href {https://doi.org/10.1103/PhysRevLett.127.046401} {\bibfield  {journal} {\bibinfo  {journal} {Physical Review Letters}\ }\textbf {\bibinfo {volume} {127}},\ \bibinfo {pages} {046401} (\bibinfo {year} {2021})}\BibitemShut {NoStop}%
\bibitem [{\citenamefont {Christensen}\ \emph {et~al.}(2021)\citenamefont {Christensen}, \citenamefont {Birol}, \citenamefont {Andersen},\ and\ \citenamefont {Fernandes}}]{Christensen2021}%
  \BibitemOpen
  \bibfield  {author} {\bibinfo {author} {\bibfnamefont {M.~H.}\ \bibnamefont {Christensen}}, \bibinfo {author} {\bibfnamefont {T.}~\bibnamefont {Birol}}, \bibinfo {author} {\bibfnamefont {B.~M.}\ \bibnamefont {Andersen}},\ and\ \bibinfo {author} {\bibfnamefont {R.~M.}\ \bibnamefont {Fernandes}},\ }\bibfield  {title} {\bibinfo {title} {Theory of the charge density wave in {$A$V$_3$Sb$_5$} kagome metals},\ }\href {https://doi.org/10.1103/PhysRevB.104.214513} {\bibfield  {journal} {\bibinfo  {journal} {Physical Review B}\ }\textbf {\bibinfo {volume} {104}},\ \bibinfo {pages} {214513} (\bibinfo {year} {2021})}\BibitemShut {NoStop}%
\bibitem [{\citenamefont {Fernandes}\ \emph {et~al.}(2019)\citenamefont {Fernandes}, \citenamefont {Orth},\ and\ \citenamefont {Schmalian}}]{Fernandes2019}%
  \BibitemOpen
  \bibfield  {author} {\bibinfo {author} {\bibfnamefont {R.~M.}\ \bibnamefont {Fernandes}}, \bibinfo {author} {\bibfnamefont {P.~P.}\ \bibnamefont {Orth}},\ and\ \bibinfo {author} {\bibfnamefont {J.}~\bibnamefont {Schmalian}},\ }\bibfield  {title} {\bibinfo {title} {Intertwined vestigial order in quantum materials: Nematicity and beyond},\ }\href {https://doi.org/10.1146/annurev-conmatphys-031218-013200} {\bibfield  {journal} {\bibinfo  {journal} {Annual Review of Condensed Matter Physics}\ }\textbf {\bibinfo {volume} {10}},\ \bibinfo {pages} {133} (\bibinfo {year} {2019})}\BibitemShut {NoStop}%
\bibitem [{\citenamefont {Tsvelik}\ and\ \citenamefont {Sarkar}(2023)}]{Tsvelik2023}%
  \BibitemOpen
  \bibfield  {author} {\bibinfo {author} {\bibfnamefont {A.~M.}\ \bibnamefont {Tsvelik}}\ and\ \bibinfo {author} {\bibfnamefont {S.}~\bibnamefont {Sarkar}},\ }\bibfield  {title} {\bibinfo {title} {Charge-density-wave fluctuation driven composite order in layered kagome metals},\ }\href {https://doi.org/10.1103/PhysRevB.108.045119} {\bibfield  {journal} {\bibinfo  {journal} {Physical Review B}\ }\textbf {\bibinfo {volume} {108}},\ \bibinfo {pages} {045119} (\bibinfo {year} {2023})}\BibitemShut {NoStop}%
\bibitem [{\citenamefont {Wildeboer}\ \emph {et~al.}(2024)\citenamefont {Wildeboer}, \citenamefont {Sarkar},\ and\ \citenamefont {Tsvelik}}]{Wildeboer2024}%
  \BibitemOpen
  \bibfield  {author} {\bibinfo {author} {\bibfnamefont {J.}~\bibnamefont {Wildeboer}}, \bibinfo {author} {\bibfnamefont {S.}~\bibnamefont {Sarkar}},\ and\ \bibinfo {author} {\bibfnamefont {A.~M.}\ \bibnamefont {Tsvelik}},\ }\bibfield  {title} {\bibinfo {title} {Phase transitions in the presence of fluctuating charge-density wave in a two-dimensional film of kagome metals},\ }\bibfield  {journal} {\bibinfo  {journal} {arXiv preprint arXiv:2411.09337}\ }\href {https://doi.org/10.48550/arXiv.2411.09337} {10.48550/arXiv.2411.09337} (\bibinfo {year} {2024})\BibitemShut {NoStop}%
\bibitem [{\citenamefont {McMillan}(1976)}]{McMillan1976}%
  \BibitemOpen
  \bibfield  {author} {\bibinfo {author} {\bibfnamefont {W.~L.}\ \bibnamefont {McMillan}},\ }\bibfield  {title} {\bibinfo {title} {Theory of discommensurations and the commensurate-incommensurate charge-density-wave phase transition},\ }\href {https://doi.org/10.1103/PhysRevB.14.1496} {\bibfield  {journal} {\bibinfo  {journal} {Physical Review B}\ }\textbf {\bibinfo {volume} {14}},\ \bibinfo {pages} {1496} (\bibinfo {year} {1976})}\BibitemShut {NoStop}%
\bibitem [{\citenamefont {Wagner}\ \emph {et~al.}(2023)\citenamefont {Wagner}, \citenamefont {Guo}, \citenamefont {Moll}, \citenamefont {Neupert},\ and\ \citenamefont {Fischer}}]{Wagner2023}%
  \BibitemOpen
  \bibfield  {author} {\bibinfo {author} {\bibfnamefont {G.}~\bibnamefont {Wagner}}, \bibinfo {author} {\bibfnamefont {C.}~\bibnamefont {Guo}}, \bibinfo {author} {\bibfnamefont {P.~J.~W.}\ \bibnamefont {Moll}}, \bibinfo {author} {\bibfnamefont {T.}~\bibnamefont {Neupert}},\ and\ \bibinfo {author} {\bibfnamefont {M.~H.}\ \bibnamefont {Fischer}},\ }\bibfield  {title} {\bibinfo {title} {Phenomenology of bond and flux orders in kagome metals},\ }\href {https://doi.org/10.1103/PhysRevB.108.125136} {\bibfield  {journal} {\bibinfo  {journal} {Physical Review B}\ }\textbf {\bibinfo {volume} {108}},\ \bibinfo {pages} {125136} (\bibinfo {year} {2023})}\BibitemShut {NoStop}%
\bibitem [{\citenamefont {Wagner}\ \emph {et~al.}(2024)\citenamefont {Wagner}, \citenamefont {Guo}, \citenamefont {Moll}, \citenamefont {Neupert},\ and\ \citenamefont {Fischer}}]{WagnerErratum2024}%
  \BibitemOpen
  \bibfield  {author} {\bibinfo {author} {\bibfnamefont {G.}~\bibnamefont {Wagner}}, \bibinfo {author} {\bibfnamefont {C.}~\bibnamefont {Guo}}, \bibinfo {author} {\bibfnamefont {P.~J.~W.}\ \bibnamefont {Moll}}, \bibinfo {author} {\bibfnamefont {T.}~\bibnamefont {Neupert}},\ and\ \bibinfo {author} {\bibfnamefont {M.~H.}\ \bibnamefont {Fischer}},\ }\bibfield  {title} {\bibinfo {title} {Erratum: Phenomenology of bond and flux orders in kagome metals [phys. rev. b 108, 125136 (2023)]},\ }\href {https://doi.org/10.1103/PhysRevB.110.159901} {\bibfield  {journal} {\bibinfo  {journal} {Physical Review B}\ }\textbf {\bibinfo {volume} {110}},\ \bibinfo {pages} {159901} (\bibinfo {year} {2024})}\BibitemShut {NoStop}%
\bibitem [{\citenamefont {Houtappel}(1950)}]{Houtappel1950}%
  \BibitemOpen
  \bibfield  {author} {\bibinfo {author} {\bibfnamefont {R.~M.~F.}\ \bibnamefont {Houtappel}},\ }\bibfield  {title} {\bibinfo {title} {Order-disorder in hexagonal lattices},\ }\href {https://doi.org/10.1016/0031-8914(50)90130-3} {\bibfield  {journal} {\bibinfo  {journal} {Physica}\ }\textbf {\bibinfo {volume} {16}},\ \bibinfo {pages} {425} (\bibinfo {year} {1950})}\BibitemShut {NoStop}%
\bibitem [{\citenamefont {Wu}(1982)}]{Wu1982}%
  \BibitemOpen
  \bibfield  {author} {\bibinfo {author} {\bibfnamefont {F.~Y.}\ \bibnamefont {Wu}},\ }\bibfield  {title} {\bibinfo {title} {The potts model},\ }\href {https://doi.org/10.1103/RevModPhys.54.235} {\bibfield  {journal} {\bibinfo  {journal} {Reviews of Modern Physics}\ }\textbf {\bibinfo {volume} {54}},\ \bibinfo {pages} {235} (\bibinfo {year} {1982})}\BibitemShut {NoStop}%
\bibitem [{\citenamefont {Iqbal}(2026)}]{Iqbal2026}%
  \BibitemOpen
  \bibfield  {author} {\bibinfo {author} {\bibfnamefont {Y.}~\bibnamefont {Iqbal}},\ }\bibfield  {title} {\bibinfo {title} {Charge order before superconductivity in the doped kagome dirac spin liquid},\ }\bibfield  {journal} {\bibinfo  {journal} {arXiv preprint arXiv:2609.15947}\ }\href {https://doi.org/10.48550/arXiv.2609.15947} {10.48550/arXiv.2609.15947} (\bibinfo {year} {2026}),\ \Eprint {https://arxiv.org/abs/2609.15947} {arXiv:2609.15947} \BibitemShut {NoStop}%
\bibitem [{\citenamefont {Yang}\ \emph {et~al.}(2020)\citenamefont {Yang}, \citenamefont {Wang}, \citenamefont {Ortiz}, \citenamefont {Liu}, \citenamefont {Gayles}, \citenamefont {Derunova}, \citenamefont {Gonzalez-Hernandez}, \citenamefont {\v{S}mejkal}, \citenamefont {Chen}, \citenamefont {Parkin}, \citenamefont {Wilson}, \citenamefont {Toberer}, \citenamefont {McQueen},\ and\ \citenamefont {Ali}}]{Yang2020}%
  \BibitemOpen
  \bibfield  {author} {\bibinfo {author} {\bibfnamefont {S.-Y.}\ \bibnamefont {Yang}}, \bibinfo {author} {\bibfnamefont {Y.}~\bibnamefont {Wang}}, \bibinfo {author} {\bibfnamefont {B.~R.}\ \bibnamefont {Ortiz}}, \bibinfo {author} {\bibfnamefont {D.}~\bibnamefont {Liu}}, \bibinfo {author} {\bibfnamefont {J.}~\bibnamefont {Gayles}}, \bibinfo {author} {\bibfnamefont {E.}~\bibnamefont {Derunova}}, \bibinfo {author} {\bibfnamefont {R.}~\bibnamefont {Gonzalez-Hernandez}}, \bibinfo {author} {\bibfnamefont {L.}~\bibnamefont {\v{S}mejkal}}, \bibinfo {author} {\bibfnamefont {Y.}~\bibnamefont {Chen}}, \bibinfo {author} {\bibfnamefont {S.~S.~P.}\ \bibnamefont {Parkin}}, \bibinfo {author} {\bibfnamefont {S.~D.}\ \bibnamefont {Wilson}}, \bibinfo {author} {\bibfnamefont {E.~S.}\ \bibnamefont {Toberer}}, \bibinfo {author} {\bibfnamefont {T.}~\bibnamefont {McQueen}},\ and\ \bibinfo {author} {\bibfnamefont {M.~N.}\ \bibnamefont {Ali}},\ }\bibfield  {title} {\bibinfo {title} {Giant, unconventional anomalous hall effect in the metallic frustrated magnet candidate {KV$_3$Sb$_5$}},\ }\href {https://doi.org/10.1126/sciadv.abb6003} {\bibfield  {journal} {\bibinfo  {journal} {Science Advances}\ }\textbf {\bibinfo {volume} {6}},\ \bibinfo {pages} {eabb6003} (\bibinfo {year} {2020})}\BibitemShut {NoStop}%
\bibitem [{\citenamefont {Xu}\ \emph {et~al.}(2022)\citenamefont {Xu}, \citenamefont {Ni}, \citenamefont {Liu}, \citenamefont {Ortiz}, \citenamefont {Deng}, \citenamefont {Wilson}, \citenamefont {Yan}, \citenamefont {Balents},\ and\ \citenamefont {Wu}}]{Xu2022}%
  \BibitemOpen
  \bibfield  {author} {\bibinfo {author} {\bibfnamefont {Y.}~\bibnamefont {Xu}}, \bibinfo {author} {\bibfnamefont {Z.}~\bibnamefont {Ni}}, \bibinfo {author} {\bibfnamefont {Y.}~\bibnamefont {Liu}}, \bibinfo {author} {\bibfnamefont {B.~R.}\ \bibnamefont {Ortiz}}, \bibinfo {author} {\bibfnamefont {Q.}~\bibnamefont {Deng}}, \bibinfo {author} {\bibfnamefont {S.~D.}\ \bibnamefont {Wilson}}, \bibinfo {author} {\bibfnamefont {B.}~\bibnamefont {Yan}}, \bibinfo {author} {\bibfnamefont {L.}~\bibnamefont {Balents}},\ and\ \bibinfo {author} {\bibfnamefont {L.}~\bibnamefont {Wu}},\ }\bibfield  {title} {\bibinfo {title} {Three-state nematicity and magneto-optical kerr effect in the charge density waves in kagome superconductors},\ }\href {https://doi.org/10.1038/s41567-022-01805-7} {\bibfield  {journal} {\bibinfo  {journal} {Nature Physics}\ }\textbf {\bibinfo {volume} {18}},\ \bibinfo {pages} {1470} (\bibinfo {year} {2022})}\BibitemShut {NoStop}%
\bibitem [{\citenamefont {Guo}\ \emph {et~al.}(2022)\citenamefont {Guo}, \citenamefont {Putzke}, \citenamefont {Konyzheva}, \citenamefont {Huang}, \citenamefont {Gutierrez-Amigo}, \citenamefont {Errea}, \citenamefont {Chen}, \citenamefont {Vergniory}, \citenamefont {Felser}, \citenamefont {Fischer}, \citenamefont {Neupert},\ and\ \citenamefont {Moll}}]{Guo2022}%
  \BibitemOpen
  \bibfield  {author} {\bibinfo {author} {\bibfnamefont {C.}~\bibnamefont {Guo}}, \bibinfo {author} {\bibfnamefont {C.}~\bibnamefont {Putzke}}, \bibinfo {author} {\bibfnamefont {S.}~\bibnamefont {Konyzheva}}, \bibinfo {author} {\bibfnamefont {X.}~\bibnamefont {Huang}}, \bibinfo {author} {\bibfnamefont {M.}~\bibnamefont {Gutierrez-Amigo}}, \bibinfo {author} {\bibfnamefont {I.}~\bibnamefont {Errea}}, \bibinfo {author} {\bibfnamefont {D.}~\bibnamefont {Chen}}, \bibinfo {author} {\bibfnamefont {M.~G.}\ \bibnamefont {Vergniory}}, \bibinfo {author} {\bibfnamefont {C.}~\bibnamefont {Felser}}, \bibinfo {author} {\bibfnamefont {M.~H.}\ \bibnamefont {Fischer}}, \bibinfo {author} {\bibfnamefont {T.}~\bibnamefont {Neupert}},\ and\ \bibinfo {author} {\bibfnamefont {P.~J.~W.}\ \bibnamefont {Moll}},\ }\bibfield  {title} {\bibinfo {title} {Switchable chiral transport in charge-ordered kagome metal {CsV$_3$Sb$_5$}},\ }\href {https://doi.org/10.1038/s41586-022-05127-9} {\bibfield  {journal} {\bibinfo  {journal} {Nature}\ }\textbf {\bibinfo {volume} {611}},\ \bibinfo {pages} {461} (\bibinfo {year} {2022})}\BibitemShut {NoStop}%
\bibitem [{\citenamefont {Saykin}\ \emph {et~al.}(2023)\citenamefont {Saykin}, \citenamefont {Farhang}, \citenamefont {Kountz}, \citenamefont {Chen}, \citenamefont {Ortiz}, \citenamefont {Shekhar}, \citenamefont {Felser}, \citenamefont {Wilson}, \citenamefont {Thomale}, \citenamefont {Xia},\ and\ \citenamefont {Kapitulnik}}]{Saykin2023}%
  \BibitemOpen
  \bibfield  {author} {\bibinfo {author} {\bibfnamefont {D.~R.}\ \bibnamefont {Saykin}}, \bibinfo {author} {\bibfnamefont {C.}~\bibnamefont {Farhang}}, \bibinfo {author} {\bibfnamefont {E.~D.}\ \bibnamefont {Kountz}}, \bibinfo {author} {\bibfnamefont {D.}~\bibnamefont {Chen}}, \bibinfo {author} {\bibfnamefont {B.~R.}\ \bibnamefont {Ortiz}}, \bibinfo {author} {\bibfnamefont {C.}~\bibnamefont {Shekhar}}, \bibinfo {author} {\bibfnamefont {C.}~\bibnamefont {Felser}}, \bibinfo {author} {\bibfnamefont {S.~D.}\ \bibnamefont {Wilson}}, \bibinfo {author} {\bibfnamefont {R.}~\bibnamefont {Thomale}}, \bibinfo {author} {\bibfnamefont {J.}~\bibnamefont {Xia}},\ and\ \bibinfo {author} {\bibfnamefont {A.}~\bibnamefont {Kapitulnik}},\ }\bibfield  {title} {\bibinfo {title} {High resolution polar kerr effect studies of {CsV$_3$Sb$_5$}: Tests for time-reversal symmetry breaking below the charge-order transition},\ }\href {https://doi.org/10.1103/PhysRevLett.131.016901} {\bibfield  {journal} {\bibinfo  {journal} {Physical Review Letters}\ }\textbf {\bibinfo {volume} {131}},\ \bibinfo {pages} {016901} (\bibinfo {year} {2023})}\BibitemShut {NoStop}%
\bibitem [{\citenamefont {Agterberg}\ and\ \citenamefont {Tsunetsugu}(2008)}]{Agterberg2008}%
  \BibitemOpen
  \bibfield  {author} {\bibinfo {author} {\bibfnamefont {D.~F.}\ \bibnamefont {Agterberg}}\ and\ \bibinfo {author} {\bibfnamefont {H.}~\bibnamefont {Tsunetsugu}},\ }\bibfield  {title} {\bibinfo {title} {Dislocations and vortices in pair-density-wave superconductors},\ }\href {https://doi.org/10.1038/nphys999} {\bibfield  {journal} {\bibinfo  {journal} {Nature Physics}\ }\textbf {\bibinfo {volume} {4}},\ \bibinfo {pages} {639} (\bibinfo {year} {2008})}\BibitemShut {NoStop}%
\bibitem [{\citenamefont {Berg}\ \emph {et~al.}(2009)\citenamefont {Berg}, \citenamefont {Fradkin},\ and\ \citenamefont {Kivelson}}]{Berg2009}%
  \BibitemOpen
  \bibfield  {author} {\bibinfo {author} {\bibfnamefont {E.}~\bibnamefont {Berg}}, \bibinfo {author} {\bibfnamefont {E.}~\bibnamefont {Fradkin}},\ and\ \bibinfo {author} {\bibfnamefont {S.~A.}\ \bibnamefont {Kivelson}},\ }\bibfield  {title} {\bibinfo {title} {Charge-$4e$ superconductivity from pair-density-wave order in certain high-temperature superconductors},\ }\href {https://doi.org/10.1038/nphys1389} {\bibfield  {journal} {\bibinfo  {journal} {Nature Physics}\ }\textbf {\bibinfo {volume} {5}},\ \bibinfo {pages} {830} (\bibinfo {year} {2009})}\BibitemShut {NoStop}%
\bibitem [{\citenamefont {Fernandes}\ and\ \citenamefont {Fu}(2021)}]{FernandesFu2021}%
  \BibitemOpen
  \bibfield  {author} {\bibinfo {author} {\bibfnamefont {R.~M.}\ \bibnamefont {Fernandes}}\ and\ \bibinfo {author} {\bibfnamefont {L.}~\bibnamefont {Fu}},\ }\bibfield  {title} {\bibinfo {title} {Charge-$4e$ superconductivity from multicomponent nematic pairing: Application to twisted bilayer graphene},\ }\href {https://doi.org/10.1103/PhysRevLett.127.047001} {\bibfield  {journal} {\bibinfo  {journal} {Physical Review Letters}\ }\textbf {\bibinfo {volume} {127}},\ \bibinfo {pages} {047001} (\bibinfo {year} {2021})}\BibitemShut {NoStop}%
\bibitem [{\citenamefont {Jian}\ \emph {et~al.}(2021)\citenamefont {Jian}, \citenamefont {Huang},\ and\ \citenamefont {Yao}}]{Jian2021}%
  \BibitemOpen
  \bibfield  {author} {\bibinfo {author} {\bibfnamefont {S.-K.}\ \bibnamefont {Jian}}, \bibinfo {author} {\bibfnamefont {Y.}~\bibnamefont {Huang}},\ and\ \bibinfo {author} {\bibfnamefont {H.}~\bibnamefont {Yao}},\ }\bibfield  {title} {\bibinfo {title} {Charge-$4e$ superconductivity from nematic superconductors in two and three dimensions},\ }\href {https://doi.org/10.1103/PhysRevLett.127.227001} {\bibfield  {journal} {\bibinfo  {journal} {Physical Review Letters}\ }\textbf {\bibinfo {volume} {127}},\ \bibinfo {pages} {227001} (\bibinfo {year} {2021})}\BibitemShut {NoStop}%
\bibitem [{\citenamefont {Lin}\ \emph {et~al.}(2025)\citenamefont {Lin}, \citenamefont {Song},\ and\ \citenamefont {Zhang}}]{Lin2025}%
  \BibitemOpen
  \bibfield  {author} {\bibinfo {author} {\bibfnamefont {T.-Y.}\ \bibnamefont {Lin}}, \bibinfo {author} {\bibfnamefont {F.-F.}\ \bibnamefont {Song}},\ and\ \bibinfo {author} {\bibfnamefont {G.-M.}\ \bibnamefont {Zhang}},\ }\bibfield  {title} {\bibinfo {title} {Theory of the charge-$6e$ condensed phase in kagome-lattice superconductors},\ }\href {https://doi.org/10.1103/PhysRevB.111.054508} {\bibfield  {journal} {\bibinfo  {journal} {Physical Review B}\ }\textbf {\bibinfo {volume} {111}},\ \bibinfo {pages} {054508} (\bibinfo {year} {2025})}\BibitemShut {NoStop}%
\bibitem [{\citenamefont {Babaev}\ \emph {et~al.}(2004)\citenamefont {Babaev}, \citenamefont {Sudbø},\ and\ \citenamefont {Ashcroft}}]{Babaev2004}%
  \BibitemOpen
  \bibfield  {author} {\bibinfo {author} {\bibfnamefont {E.}~\bibnamefont {Babaev}}, \bibinfo {author} {\bibfnamefont {A.}~\bibnamefont {Sudbø}},\ and\ \bibinfo {author} {\bibfnamefont {N.~W.}\ \bibnamefont {Ashcroft}},\ }\bibfield  {title} {\bibinfo {title} {A superconductor to superfluid phase transition in liquid metallic hydrogen},\ }\href {https://doi.org/10.1038/nature02910} {\bibfield  {journal} {\bibinfo  {journal} {Nature}\ }\textbf {\bibinfo {volume} {431}},\ \bibinfo {pages} {666–668} (\bibinfo {year} {2004})}\BibitemShut {NoStop}%
\bibitem [{\citenamefont {Herland}\ \emph {et~al.}(2010)\citenamefont {Herland}, \citenamefont {Babaev},\ and\ \citenamefont {Sudb\o{}}}]{Herland2010}%
  \BibitemOpen
  \bibfield  {author} {\bibinfo {author} {\bibfnamefont {E.~V.}\ \bibnamefont {Herland}}, \bibinfo {author} {\bibfnamefont {E.}~\bibnamefont {Babaev}},\ and\ \bibinfo {author} {\bibfnamefont {A.}~\bibnamefont {Sudb\o{}}},\ }\bibfield  {title} {\bibinfo {title} {Phase transitions in a three dimensional $u(1)\ifmmode\times\else\texttimes\fi{}u(1)$ lattice london superconductor: Metallic superfluid and charge-$4e$ superconducting states},\ }\href {https://doi.org/10.1103/PhysRevB.82.134511} {\bibfield  {journal} {\bibinfo  {journal} {Phys. Rev. B}\ }\textbf {\bibinfo {volume} {82}},\ \bibinfo {pages} {134511} (\bibinfo {year} {2010})}\BibitemShut {NoStop}%
\bibitem [{\citenamefont {Shi}(2026)}]{shi_2026_23254801}%
  \BibitemOpen
  \bibfield  {author} {\bibinfo {author} {\bibfnamefont {Y.}~\bibnamefont {Shi}},\ }\href {https://doi.org/10.5281/zenodo.23254801} {\bibinfo {title} {Monte carlo code for vestigial chirality from fluctuating loop currents on the kagome lattice}} (\bibinfo {year} {2026})\BibitemShut {NoStop}%
\bibitem [{\citenamefont {Swendsen}\ and\ \citenamefont {Wang}(1987)}]{SwendsenWang1987}%
  \BibitemOpen
  \bibfield  {author} {\bibinfo {author} {\bibfnamefont {R.~H.}\ \bibnamefont {Swendsen}}\ and\ \bibinfo {author} {\bibfnamefont {J.-S.}\ \bibnamefont {Wang}},\ }\bibfield  {title} {\bibinfo {title} {Nonuniversal critical dynamics in monte carlo simulations},\ }\href {https://doi.org/10.1103/PhysRevLett.58.86} {\bibfield  {journal} {\bibinfo  {journal} {Physical Review Letters}\ }\textbf {\bibinfo {volume} {58}},\ \bibinfo {pages} {86} (\bibinfo {year} {1987})}\BibitemShut {NoStop}%
\bibitem [{\citenamefont {Hukushima}\ and\ \citenamefont {Nemoto}(1996)}]{HukushimaNemoto1996}%
  \BibitemOpen
  \bibfield  {author} {\bibinfo {author} {\bibfnamefont {K.}~\bibnamefont {Hukushima}}\ and\ \bibinfo {author} {\bibfnamefont {K.}~\bibnamefont {Nemoto}},\ }\bibfield  {title} {\bibinfo {title} {Exchange monte carlo method and application to spin glass simulations},\ }\href {https://doi.org/10.1143/JPSJ.65.1604} {\bibfield  {journal} {\bibinfo  {journal} {Journal of the Physical Society of Japan}\ }\textbf {\bibinfo {volume} {65}},\ \bibinfo {pages} {1604} (\bibinfo {year} {1996})}\BibitemShut {NoStop}%
\bibitem [{\citenamefont {Maccari}\ \emph {et~al.}(2023)\citenamefont {Maccari}, \citenamefont {Carlstr\"om},\ and\ \citenamefont {Babaev}}]{Maccari2023}%
  \BibitemOpen
  \bibfield  {author} {\bibinfo {author} {\bibfnamefont {I.}~\bibnamefont {Maccari}}, \bibinfo {author} {\bibfnamefont {J.}~\bibnamefont {Carlstr\"om}},\ and\ \bibinfo {author} {\bibfnamefont {E.}~\bibnamefont {Babaev}},\ }\bibfield  {title} {\bibinfo {title} {Prediction of time-reversal-symmetry breaking fermionic quadrupling condensate in twisted bilayer graphene},\ }\href {https://doi.org/10.1103/PhysRevB.107.064501} {\bibfield  {journal} {\bibinfo  {journal} {Physical Review B}\ }\textbf {\bibinfo {volume} {107}},\ \bibinfo {pages} {064501} (\bibinfo {year} {2023})}\BibitemShut {NoStop}%
\bibitem [{\citenamefont {Vehtari}\ \emph {et~al.}(2021)\citenamefont {Vehtari}, \citenamefont {Gelman}, \citenamefont {Simpson}, \citenamefont {Carpenter},\ and\ \citenamefont {B\"urkner}}]{Vehtari2021}%
  \BibitemOpen
  \bibfield  {author} {\bibinfo {author} {\bibfnamefont {A.}~\bibnamefont {Vehtari}}, \bibinfo {author} {\bibfnamefont {A.}~\bibnamefont {Gelman}}, \bibinfo {author} {\bibfnamefont {D.}~\bibnamefont {Simpson}}, \bibinfo {author} {\bibfnamefont {B.}~\bibnamefont {Carpenter}},\ and\ \bibinfo {author} {\bibfnamefont {P.-C.}\ \bibnamefont {B\"urkner}},\ }\bibfield  {title} {\bibinfo {title} {Rank-normalization, folding, and localization: An improved {$\widehat R$} for assessing convergence of {MCMC}},\ }\href {https://doi.org/10.1214/20-BA1221} {\bibfield  {journal} {\bibinfo  {journal} {Bayesian Analysis}\ }\textbf {\bibinfo {volume} {16}},\ \bibinfo {pages} {667} (\bibinfo {year} {2021})}\BibitemShut {NoStop}%
\bibitem [{\citenamefont {Ferrenberg}\ and\ \citenamefont {Swendsen}(1989)}]{FerrenbergSwendsen1989}%
  \BibitemOpen
  \bibfield  {author} {\bibinfo {author} {\bibfnamefont {A.~M.}\ \bibnamefont {Ferrenberg}}\ and\ \bibinfo {author} {\bibfnamefont {R.~H.}\ \bibnamefont {Swendsen}},\ }\bibfield  {title} {\bibinfo {title} {Optimized monte carlo data analysis},\ }\href {https://doi.org/10.1103/PhysRevLett.63.1195} {\bibfield  {journal} {\bibinfo  {journal} {Physical Review Letters}\ }\textbf {\bibinfo {volume} {63}},\ \bibinfo {pages} {1195} (\bibinfo {year} {1989})}\BibitemShut {NoStop}%
\end{thebibliography}%

\end{document}


\title{Supplementary Information for ``Vestigial chirality from
fluctuating loop currents on the kagome lattice''}

\author{Yin Shi}
\email{yin.shi@iphy.ac.cn}
\affiliation{Beijing National Laboratory for Condensed Matter Physics and Institute of Physics, Chinese Academy of Sciences, Beijing 100190, China}%
\author{Xuande Bu}
\affiliation{Beijing National Laboratory for Condensed Matter Physics and Institute of Physics, Chinese Academy of Sciences, Beijing 100190, China}%
\affiliation{University of Chinese Academy of Sciences, Beijing 100049, China}%
\author{Sheng Meng}
\email{smeng@iphy.ac.cn}
\affiliation{Beijing National Laboratory for Condensed Matter Physics and Institute of Physics, Chinese Academy of Sciences, Beijing 100190, China}%
\affiliation{University of Chinese Academy of Sciences, Beijing 100049, China}%
\affiliation{Songshan Lake Materials Laboratory, Dongguan, Guangdong 523808, China}%

\date{\today}

\maketitle

\section{Transition summaries}

Tables~\ref{tab:sign-crossings} and \ref{tab:soft-splits} collect the numerical
transition estimates quoted in the main text.  They are separated from the
main article to keep its Results focused on the phase topology and physical
mechanism.

\begin{table}[!htbp]
\caption{Representative sign-model crossings.  Temperatures are in units of
$K_1$.  Parentheses denote one-standard-error uncertainty in the last shown
digits; grid-scale floors are included where needed.}
\label{tab:sign-crossings}
\begin{ruledtabular}
\begin{tabular}{c c c c}
$q$ & $T_s$ & $T_\tau$ & interpretation\\
\hline
0.00 & 3.634(20) & 3.634(20) & direct\\
1.00 & 5.080(5)  & 5.080(5)  & direct\\
1.24 & 5.382(3)  & 5.387(3)  & unresolved\\
1.26 & 5.403(3)  & 5.414(3)  & first resolved\\
1.28 & 5.420(3)  & 5.445(3)  & vestigial\\
2.00 & 5.7131(7) & 7.4592(8) & vestigial\\
3.00 & 5.751(20) & 10.970(6) & vestigial\\
\end{tabular}
\end{ruledtabular}
\end{table}

\begin{table}[!htbp]
\caption{Exploratory $L=12$--$36$ joint-block-bootstrap estimates of
$({\Tchi-\TPhi})/K_\Phi$ in the amplitude-resolved onset scan.  Each column
labels a successive pair of sizes; the promoted $J_\chi/K_\Phi=5.5$ large-size sequence is
listed separately in Table~\ref{tab:soft-large}.}
\label{tab:soft-splits}
\begin{ruledtabular}
\begin{tabular}{c c c c c c}
$J_\chi/K_\Phi$ & $12$--$18$ & $18$--$24$ & $24$--$30$ & $30$--$36$ & conclusion\\
\hline
4.75 & $0.00028(83)$ & $0.00370(130)$ & $-0.00034(62)$ & $0.00096(62)$
     & direct/crossover\\
5.00 & $0.00318(112)$ & $0.00370(86)$ & $0.00349(121)$ & $0.00188(72)$
     & thermodynamic limit unresolved\\
5.50 & $0.02339(234)$ & $0.02056(303)$ & $0.02451(386)$ & $0.02422(454)$
     & small-size split\\
\end{tabular}
\end{ruledtabular}
\end{table}

\begin{table}[!htbp]
\caption{Promoted three-field $J_\chi/K_\Phi=5.5$ crossing sequence from the
second-moment ratios.  Uncertainties are joint chain-and-block bootstrap
standard errors.}
\label{tab:soft-large}
\begin{ruledtabular}
\begin{tabular}{c c c c}
size pair & $T_\Phi/K_\Phi$ & $T_\chi/K_\Phi$ & $(T_\chi-T_\Phi)/K_\Phi$\\
\hline
$24$--$48$  & $2.7695(4)$  & $2.8000(18)$ & $0.0306(18)$\\
$48$--$96$  & $2.76975(16)$ & $2.81318(71)$ & $0.04343(75)$\\
$96$--$144$ & $2.76973(16)$ & $2.81370(40)$ & $0.04396(43)$\\
$144$--$192$ & $2.76918(25)$ & $2.81310(64)$ & $0.04392(65)$\\
\end{tabular}
\end{ruledtabular}
\end{table}

\FloatBarrier

\section{Sign-model numerical details}

\subsection{Refined transitions at \texorpdfstring{$q=2$}{q=2}}

For the lower boundary, eight independent parallel-tempering ladders were
simulated at each of $L=96,192,288,384$ on a common grid
$5.68\leq T/K_1\leq5.78$.  At least 4096 measurements were retained per
temperature and chain.  All sizes satisfy split-$\hat R<1.05$; the $L=384$
ensemble records 15 aggregate cold--hot--cold round trips, with repeated
bidirectional passages in every chain.  Joint block bootstrap gives successive
$R_s$ crossings $5.71298(21)$, $5.71226(25)$, and $5.71306(71)$.  No stable
two-peak energy histogram is present.

For the upper boundary, eight independent ladders at each size were run on a
33-slot grid $7.40\leq T/K_1\leq7.52$, retaining 8192 measurements per
temperature and chain.  Every primitive-moment series has
split-$\hat R<1.009$; the $L=384$ ensemble records 2654 aggregate round trips.
The successive $R_\tau$ crossings are $7.4580(7)$, $7.4588(7)$, and
$7.4592(8)$.  Binder crossings agree within uncertainty, and the energy
histograms again show no two-peak candidate.

The real-space correlation data combine four chains and 16 blocks per chain.
Four measurements per block are separated by four embedded-cluster sweeps.
Uncertainties are the standard errors across chain means.

\subsection{Endpoint criterion and histogram checks}

The endpoint corridor is transport limited: replicas can accept many
neighboring exchanges without traversing the full free-energy bottleneck.  We
therefore classify a split as resolved only when
\begin{equation}
 \Delta T=T_\tau-T_s>
 2\left(\sigma_s^2+\sigma_\tau^2\right)^{1/2}.
 \label{eq:supp-endpoint-criterion}
\end{equation}
Here $\sigma_s$ and $\sigma_\tau$ are the one-standard-error uncertainties of
$T_s$ and $T_\tau$, respectively.
By this criterion, the split is unresolved at $q=1.24$ and first resolved, at
the native-grid precision, at $q=1.26$.  The entries in Table~S1 use the
independent $L=192$--$288$ feedback-tempering campaign,
which gives $(T_s,T_\tau)/K_1=(5.3823,5.3871)$ at $q=1.24$ and
$(5.4030,5.4145)$ at $q=1.26$.  No large-size ladder completes a full
cold--hot--cold round trip, and the $q=1.24$ primary estimator has
split-$\hat R=1.144$.  We therefore retain the half-grid uncertainty floor and
do not quote a more precise value of $q_c$.

Energy-histogram scaling does not establish that the direct line is first
order.  The largest-size distributions are predominantly single peaked, and
apparent equal-weight barriers at transport-limited points are not stable
against size and chain checks.  Determining whether the direct line is
continuous or weakly first order is distinct from identifying where the two
thermal boundaries separate.

\subsection{Finite-size tails above the transitions}

At $T/K_1=7.69>T_\tau/K_1$, whole-chain bootstrap fits over
$L=48,96,192$ give
\begin{equation}
 \sqrt{\langle m_s^2\rangle}\propto L^{-1.012(8)},\qquad
 \sqrt{\langle m_\tau^2\rangle}\propto L^{-0.950(29)}.
\end{equation}
Both are consistent with the disordered two-dimensional law $L^{-1}$.  At
fixed $L=96$, the chiral root-mean-square amplitude decreases from
$0.2052(26)$ at $T/K_1=7.69$ to $0.09873(95)$ at $8.0$ and $0.05990(58)$ at
$8.5$.  The apparent high-temperature tails in the main-text thermal curves
are therefore finite-size susceptibility backgrounds rather than residual
order.

\FloatBarrier

\section{Amplitude-resolved numerical details}

\subsection{Parameter family and stability}

We take $K_\Phi$ as the energy unit and restore amplitude fluctuations using
\begin{align}
 \frac{r_\Phi}{K_\Phi}&=-400,&
 \frac{u_\Phi}{K_\Phi}&=400,&
 \frac{v_\Phi}{K_\Phi}&=440,\nonumber\\
 \frac{c_\Phi}{K_\Phi}&=100,&
 \frac{J_\chi}{K_\Phi}&=4.75,5.0,5.5.
 \label{eq:supp-soft-parameters}
\end{align}
The common scale of the local coefficients is deliberately large:
\begin{equation}
 \frac{(r_\Phi,u_\Phi,v_\Phi,c_\Phi)}{K_\Phi}
 =100(-4,4,4.4,1).
\end{equation}
These are the cross-component-basis coefficients of the main-text Hamiltonian.
Equivalently, the isotropic-plus-cubic-anisotropy convention used by the
simulation input has coefficients
$\tilde u_\Phi=u_\Phi-v_\Phi=-40K_\Phi$ and
$\tilde v_\Phi=v_\Phi=440K_\Phi$.
The quartic potential is stable both along an equal-component direction, where
$(u_\Phi-2v_\Phi/3)/K_\Phi=106.67$, and along a one-component direction, where
$u_\Phi/K_\Phi=400$.  For a uniform equal-amplitude state
$\Phi_a=A$, the normalized energy per site is
\begin{equation}
 \frac{V_{\rm unif}(A)}{K_\Phi}
 =-600A^2+240A^4+
 \left(450-3\frac{J_\chi}{K_\Phi}\right)A^6.
 \label{eq:supp-uniform-potential}
\end{equation}
At $J_\chi/K_\Phi=5.5$, its zero-temperature minimum is $A_0=0.721$, close to the
measured transition-region root-mean-square amplitude $0.714$.  This stiff,
bounded family is a controlled bridge from the sign limit rather than a
material fit.

\subsection{Simulation campaign and convergence}

The exploratory $L=12$--$36$ runs used a general six-field (three bond fields plus three loop-current fields) implementation
whose additional massive block was exactly decoupled from the three loop-current
fields.  A dedicated three-field implementation reproduces the archived
loop-current observables on the same temperature ladder and removes the
spectator from updates, reductions, and checkpoints.  The promoted
$J_\chi/K_\Phi=5.5$ campaign uses $L=24,48,96,144,192$, four independent
chains, and a common 65-temperature feedback ladder over
$2.68\leq T/K_\Phi\leq2.84$.  Exchanges
are attempted every two sweeps, and each chain-temperature pair retains 2048
measurements separated by ten sweeps.

All five sizes pass the preregistered convergence gates.  From smallest to
largest, the maximum split-$\hat R$ values are $1.0015$, $1.0011$, $1.0022$,
$1.0057$, and $1.0044$; the minimum effective sample sizes are $6648$, $5230$,
$3320$, $1509$, and $697$.  The $L=192$ run has minimum neighboring acceptance
$0.414$, a longest autocorrelation time of $71.2$ sweeps, and 26 distinct
walkers that complete a cold--hot--cold round trip.

\subsection{Crossings and order-content test}

At $J_\chi/K_\Phi=4.75$, the small-size crossing sequence changes sign and
belongs to the direct/crossover side.  At $J_\chi/K_\Phi=5.0$, reasonable
correction forms disagree, so that case remains unresolved.  For
$J_\chi/K_\Phi=5.5$, the $\xi/L$ sequence in Table~\ref{tab:soft-large}
converges to the largest-pair values
$(\TPhi/K_\Phi,\Tchi/K_\Phi)=[2.76918(25),2.81310(64)]$, with normalized
separation $(\Tchi-\TPhi)/K_\Phi=0.04392(65)$.  Correlation-ratio crossings
give the consistent separation $0.04358(43)$; all 4000 joint chain-and-block
bootstrap draws have positive separation.

At $T/K_\Phi=2.79121$, inside the crossing interval, the normalized primary structure
factor is
\begin{equation}
 \frac{S_\Phi(0)}{L^2}=
 0.4406,\ 0.1129,\ 0.02340,\ 0.01032,\ 0.005767,
 \label{eq:supp-Sphi-scaling}
\end{equation}
for $L=24,48,96,144,192$, whereas
\begin{equation}
 \frac{S_\chi(0)}{L^2}=
 0.08256,\ 0.07061,\ 0.06818,\ 0.06800,\ 0.06770.
 \label{eq:supp-Schi-scaling}
\end{equation}
Power-law descriptions over the simulated sizes give
$S_\Phi(0)/L^2\sim L^{-2.11}$ and
$S_\chi(0)/L^2\sim L^{-0.09}$.  The primary structure factor is consistent
with saturation at a finite short-range value, while the time-reversal-odd
composite remains extensive.

\subsection{Matching to the rigid limit}

The bulk-amplitude estimate
\begin{equation}
 q_\chi^{\rm eff}(T)=\frac{J_\chi}{K_\Phi}
 \left[\frac{\langle\Phi_1^2+\Phi_2^2+\Phi_3^2\rangle_T}{3}\right]^2
 \label{eq:supp-qeff}
\end{equation}
gives $q_\chi^{\rm eff}\simeq1.23,1.30,1.43$ for the three focused couplings.
Because a wall can suppress the amplitude in its core, this bulk estimate is
only a guide: the chiral scale varies as $A^6$ while the primary stiffness
varies as $A^2$.  Fully relaxed wall tensions are the appropriate quantitative
matching variables.

\FloatBarrier

\section{Binder-ratio consistency}

\begin{figure}[t]
 \centering
 \includegraphics[width=0.96\textwidth]{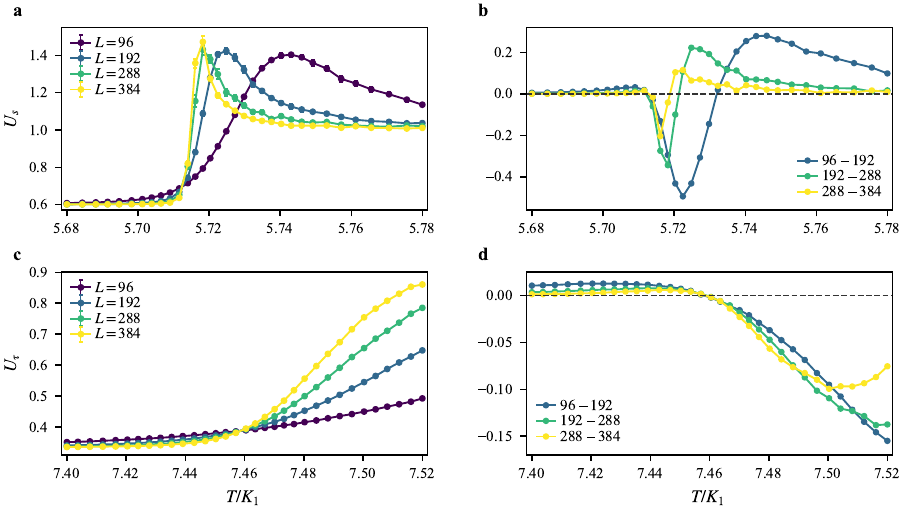}
 \caption{Finite-size Binder consistency diagnostics at $q=2$.
 (a) Primary Binder ratio $U_s(T)$ on the common temperature grid of main-text
 Fig.~2b.  (b) Successive-size differences have two zero crossings:
 $(5.71254,5.73202)$, $(5.71211,5.72137)$, and $(5.71280,5.71884)$ for the
 $96$--$192$, $192$--$288$, and $288$--$384$ pairs, respectively.
 (c) Chiral Binder ratio $U_\tau(T)$ on the grid of main-text Fig.~2c.
 (d) Its unique successive-size crossings are $7.4582(9)$, $7.4591(8)$, and
 $7.4587(9)$, consistent with the $R_\tau$ sequence $7.4580(7)$, $7.4588(7)$,
 and $7.4592(8)$.  Error bars in panels (a,c) are from the joint
 primitive-moment bootstrap.}
 \label{fig:binder}
\end{figure}

The dedicated $q=2$ campaigns also measure the Binder ratios
\begin{equation}
 U_s=\frac{3\langle m_s^4\rangle}{5\langle m_s^2\rangle^2},
 \qquad
 U_\tau=\frac{\langle m_\tau^4\rangle}{3\langle m_\tau^2\rangle^2}.
\end{equation}
Figure~\ref{fig:binder}(a) shows that the primary curves develop a pronounced
nonmonotonic maximum.  Every successive-size difference in panel (b)
consequently changes sign twice.  The lower roots track the unique $R_s$
crossing used in the main text, whereas the upper roots drift strongly with
size.  This two-root ambiguity is why the primary Binder ratio is retained as
a consistency diagnostic rather than used as the transition locator.

The chiral Binder curves in Fig.~\ref{fig:binder}(c) are monotonic over the
refined upper-transition window.  Their successive-size differences in panel
(d) have one zero each, providing an unambiguous independent check of the
main-text $R_\tau$ crossings.